\documentclass[10pt,twocolumn,twoside] {IEEEtran}
\usepackage{amsmath}
\usepackage{cite,url}
\usepackage{amsfonts}
\usepackage{fancyhdr}
\usepackage{amsmath,amssymb,amsthm}%
\usepackage{graphicx,psfrag,caption,subcaption}%
\usepackage{listings}%
\usepackage{arydshln}
\usepackage{subfig}%
\usepackage{algorithm,algorithmic}
\newtheorem{mytheo}{Theorem}
\newtheorem*{myproblem}{Problem statement}%
\newtheorem{myrem}{Remark}%

\renewcommand{\thefootnote}{\fnsymbol{footnote}}
\DeclareMathOperator*{\argmin}{arg\,min}
\DeclareMathOperator*{\argmax}{arg\,max}

\begin{document}

\title{Sparsity-Promoting Sensor Selection for \\Non-linear Measurement Models}
\author{Sundeep Prabhakar Chepuri,~\IEEEmembership{Student Member,~IEEE,} and Geert~Leus,~\IEEEmembership{Fellow,~IEEE} 
                                
\thanks{This work was supported in part by STW under the FASTCOM project (10551) and in part by NWO-STW under the VICI program (10382).}

\thanks{All the authors are with the Faculty of Electrical Engineering, Mathematics and Computer Science, Delft University of Technology, The Netherlands. 
Email:~\{s.p.chepuri;g.j.t.leus\}@tudelft.nl.}
\thanks{A conference precursor of this manuscript has been
published in~\cite{Eusipco13chepuri}.}
}
\maketitle
\thispagestyle{empty}
\pagenumbering{arabic}
\renewcommand{\thefootnote}{\arabic{footnote}}
\IEEEoverridecommandlockouts
\maketitle
\begin{abstract}
Sensor selection is an important design problem in large-scale sensor networks. Sensor selection can be interpreted as the problem of selecting the best subset of sensors that guarantees a certain estimation performance. We focus on observations that are related to a general non-linear model. The proposed framework is valid as long as the observations are independent, and its likelihood satisfies the regularity conditions. We use several functions of the Cram\'er-Rao bound (CRB) as a performance measure. We formulate the sensor selection problem as the design of a selection vector, which in its original form is a nonconvex $\ell_0$-(quasi) norm optimization problem. We present relaxed sensor selection solvers that can be efficiently solved in polynomial time. We also propose a projected subgradient algorithm that is attractive for large-scale problems and also show how the algorithm can be easily distributed. The proposed framework is illustrated with a number of examples related to sensor placement design for localization.
\end{abstract}
\begin{keywords}
Sensor selection, sensor placement, Cram\'er-Rao bound, selection vector, sparsity, non-linear models, statistical inference, projected subgradient algorithm, convex optimization, sensor networks.
\end{keywords}
%
\section{Introduction}
\IEEEPARstart{A}{}dvances in sensor technology have enabled a large spectrum of applications and services related to safety and security,  surveillance, environmental and climate monitoring, to list a few. The sensor nodes are spatially deployed and operate as a network, with each sensor node capable of sensing, processing, and communicating to other nodes or a central processing unit. As a network, their fundamental task is distributed data sampling (i.e., to sense the environment) from which we seek to extract relevant information. The sensors provide a prohibitively large dataset which is usually gathered at a fusion center. This gathered data has to be optimally processed, rejecting the redundant, identical, or faulty measurements. 

Sensor selection is a fundamental design task in sensor networks. The number of sensors are often limited either by economical constraints (hardware costs), or the availability of physical or storage space. In order to reduce the hardware costs, as well as the resulting communications and processing overhead, one would like to smartly deploy the sensors. Sensor selection also enables the design of spatio-temporal sensing patterns that guarantee a certain performance measure such as energy-efficiency, information measure, estimation accuracy, or detection probability. The sensor placement problem can also be interpreted as a sensor selection problem in which the best subset of the available sensor locations are selected subject to a specific performance constraint. Sensor selection is pertinent to various diverse fields, especially to applications dealing with large-scale networks like network monitoring~\cite{NetworkCartography,giannakis2013monitoring}, location-aware services like target localization and tracking~\cite{Gusta05SPM,GinnakisLoc,localizationSPM}, field estimation~\cite{ToonField,MouraField}, and environment monitoring in general. The fundamental questions of interest are: 
\begin{enumerate}
\item[q1.] Where to deploy the limited sensors available?
\item[q2.] Do we need to process all the acquired measurements?
\end{enumerate}
To this end, we focus on processing only the most informative sensors  for a general {\it non-linear} statistical inference problem. 

\subsection{Related prior works}

A large volume of literature exists on sensor selection~\cite[and references therein]{sensel1}. The sensor selection problem is often formulated as an optimization problem based on some well-known performance measures from experimental design~\cite{pukelsheim1993optimal,sensel1},~\cite[Pg. 384]{Boyd}. The sensor selection problem is expressed as the following optimization problem:
\begin{equation}
\label{eq:design}
\argmin_{{\bf w} \in \{0,1\}^M} \quad f({\bf E}({\bf w})) \quad {\rm s.t.} \quad {\bf 1}_M^T{\bf w} = K,
\end{equation}
where ${\bf w}$ is a selection vector of length $M$, and $f({\bf E}({\bf w}))$ is a scalar cost function related to the mean squared error (MSE) covariance matrix ${\bf E}$. The MSE covariance matrix is optimized to select the best subset of $K$ sensors out of $M$ available sensors such that $K \ll M$. Different functions $f({\bf E}({\bf w}))$ can be used, and the typical choices for $f({\bf E}({\bf w}))$ are related to:
\begin{description}
\item[1.]{\it A-optimality}: minimizes the sum of eigenvalues of  ${\bf E}$ with $f({\bf w}):= {\rm tr}\{{\bf E}({\bf w})\}.$
\item[2.]{\it E-optimality}: minimizes the maximum eigenvalue of ${\bf E}$ with $f({\bf w}):= \lambda_{\max}\{{\bf E}({\bf w})\}.$
\item[3.]{\it D-optimality}: minimizes the determinant of ${\bf E}$ with $f({\bf w}):= \ln {\rm det}\{{\bf E}({\bf w})\}.$
\end{description}
This is a combinatorial optimization problem involving $\binom{K}{M}$ searches, and it is clearly intractable even for small-scale problems with $K=10$ and $M=100$. To simplify this problem, the nonconvex Boolean constraint ${\bf w} \in \{0,1\}^M$ is relaxed to a convex box constraint ${\bf w} \in [0,1]^M$. The relaxed optimization problem has been studied in~\cite{sensel1} for additive Gaussian linear models, where the matrix ${\bf E}$ is available in closed form, and more importantly, where the above listed performance measures are {\it independent} of the unknown parameter. Moreover, in practice, the exact number of sensors ${K}$ to select might not be known. However, this number $K$ can always be tuned to achieve a desired performance.

The above selection problem is applied to sensor placement for power grid monitoring in~\cite{PMUplacementGG2012}. 
Alternative approaches exploiting the submodularity of the objective function~\cite{krause2008submodularity,krause2007near,krause2008robust}, heuristics based on genetic algorithms~\cite{geneticselection}, and greedy algorithms~\cite{shamaiah2010greedy} are also proposed to solve the sensor selection problem. Sensor selection for dynamical systems often referred to as sensor polling or scheduling, is studied in~\cite{varshneySPL,Carmischeduling,GerrySensormanagement2012}. In~\cite{vertterlisensorplacement}, the  sensor placement problem for linear models is addressed as the design of a sensing matrix that optimizes a measure related to the orthogonality of its rows. All the above literature (in general) deals with measurements that are related to additive Gaussian linear models. Experimental design for non-linear models within the Bayesian and sequential design frameworks is discussed in~\cite{ExptdesignNon}. In~\cite{varshneySPL}, sensor selection for target tracking based on extended Kalman filtering (EKF) has been proposed, in which the selection is performed by designing an appropriate gain matrix. Although a non-linear measurement model in additive Gaussian noise is used in~\cite{varshneySPL}, the past state estimate (not the true state) is used to compute the error covariance matrix leading to a suboptimal solution. Sensor selection for detection problems is studied in~\cite{SenselDetection}. In~\cite{kekatos2011sparse}, reliable sensor selection based on the actual measurements to identify the outliers is presented. A different problem, yet related to sensor selection, is the problem of identifying source-informative sensors, which is studied in~\cite{Schizas2013TSP}.

\subsection{Contributions}
The {\it sensor selection} problem can be interpreted as the problem to select the best sensors out of $M$ available sensors. The selected sensors are deemed as the best subset of sensors if they guarantee a certain specified estimation accuracy. We consider general scenarios where the measurements of the unknown parameter follow a {\it non-linear model} (unlike~\cite{sensel1} for instance). Non-linear measurement models are frequently encountered in applications like source localization, tracking, field estimation, or phase retrieval, to list a few. The error covariance matrix for non-linear models is not always available in closed form, and more importantly it depends on the unknown parameter. Our first contribution in the context of sensor selection is to use the {\it Cram\'er-Rao bound} (CRB) as a performance measure.
The CRB is a rigorous performance measure for optimality, and it generalizes very well for non-linear measurement models (not necessarily in additive Gaussian noise). Moreover, we do not need the actual measurements, and hence, our framework is also well-suited for solving offline design problems. In addition to this, the number of sensors that have to be selected, i.e., $K$, is generally not known in practice. Hence, instead of fixing $K$ as in \eqref{eq:design}, we pose sensor selection as a cardinality minimization problem that provides the number of selected sensors as a byproduct. In order to do this, we use different thresholds that specify the required accuracy. 

The proposed sensor selection framework is very generic and can be applied to any non-linear statistical inference problem (linear being a special case). The selection problem is formulated as the design of a {\it selection} vector which is an~$\ell_0$-(quasi) norm nonconvex {\it Boolean} optimization problem. It requires a brute-force evaluation over all the $2^M$ choices. For example, with $M=100$ available potential sensors, there are in the order of $10^{30}$ possible choices whose direct enumeration is clearly impossible.
The nonconvex sensor selection problem is relaxed using standard convex relaxation techniques which can then be efficiently solved in polynomial time. 

To cope with large-scale problems, we further present a {\it projected subgradient} algorithm. It is worth mentioning that the projected subgradient algorithm allows a very easy distributed implementation. 

A {\it sparsity-enhancing} concave surrogate for the $\ell_0$-(quasi) norm is also proposed for sensor selection as an alternative to the traditional best convex relaxation. This is particularly advantageous when there are multiple (nearly) identical sensor measurements. We illustrate the sensor selection problem using examples of sensor placement for source localization. 

\subsection{Outline and notations}
The remainder of the paper is organized as follows. In Section~\ref{sec:model}, we present the non-linear measurement model. In Section~\ref{sec:probfor}, we show the problem formulation, and we present the algorithms that solve the relaxed optimization problem in Section~\ref{sec:optprob}.
In Section~\ref{sec:extension}, we derive the dual problem, and provide some extensions. In Section~\ref{sec:example}, the proposed framework is applied to a number of different models related to sensor selection for localization. The paper finally concludes with Section~\ref{sec:conclusion}.

The notations used in this paper can be described as follows. Upper (lower) bold face letters are used for matrices (column vectors). $(\cdot)^T$ denotes transposition. $\mathrm{diag}(\cdot)$ refers to a block diagonal matrix with the elements in its argument on the main diagonal. $\mathbf{1}_N$ ($\mathbf{0}_N$) denotes the $N \times 1$ vector of ones (zeros). $\mathbf{I}_N$ is an identity matrix of size $N$. $\mathbb{E}\{\cdot\}$ denotes the expectation operation. ${\rm tr}\{\cdot\}$ is the matrix trace operator. ${\rm det}\{\cdot\}$ is the matrix determinant. $\lambda_{\rm min}\{{\bf A}\}$ ($\lambda_{\rm max}\{{\bf A}\}$) denotes the minimum (maximum) eigenvalue of a symmetric matrix ${\bf A}$. ${\bf A} \succeq {\bf B}$ means that ${\bf A} - {\bf B}$ is a positive semidefinite matrix. $\mathbb{S}^{N}$ ($\mathbb{S}^{N}_{+}$) denotes the set of symmetric (symmetric positive semi-definite) matrices of size $N \times N$. $|\mathcal{U}|$ denotes the cardinality of the set $\mathcal{U}$.  
\section{Non-linear measurement model}\label{sec:model}
In this paper, we consider a generic non-linear measurement model 
\begin{equation}
\label{eq:genericmodel}
{y}_m = {h}_m ({\boldsymbol \theta},{n}_m), \, m=1,2,\ldots,M, 
\end{equation}
where ${y}_m$ is the $m$th spatial or temporal sensor measurement, ${\boldsymbol \theta} \in \mathbb{R}^N$ is the unknown parameter, 
${n}_m$ for $m=1,2,\ldots,M,$ is the noise process, and the regressors $h_m$ for $m=1,2,\ldots,M,$ are (in general) non-linear functionals. 
Let the vector ${\bf y} = [y_1,y_2,\ldots,y_M]^T \in \mathbb{R}^M$ collect the measurements. The likelihood of the measurements $p({\bf y};{\boldsymbol \theta})$ is the probability density function (pdf) of ${\bf y}$ parameterized by the unknown vector ${\boldsymbol \theta}$. 

We make the following assumptions:
\begin{description}
\item[a1.]{\it Regularity conditions}: The log-likelihood of the measurements satisfies the regularity condition $\mathbb{E}\{\frac{\partial \ln p({\bf y}; {\boldsymbol \theta})}{\partial {\boldsymbol \theta}}\} = 0$. This is a well-known condition for the CRB to exist~\cite{SKayestimation}.
\item[a2.]{\it Independent observations}: The measurements $y_m$ for $m=1,2,\ldots,M,$ are a sequence of independent random variables. 
\end{description}
The proposed framework for sensor selection is valid as long as the above two assumptions hold. 

Assuming (a1) holds, the covariance of any unbiased estimate $\hat{\boldsymbol \theta} \in \mathbb{R}^N$ of the unknown parameter satisfies the well-known inequality~\cite{SKayestimation}
\begin{equation*}
\label{eq:CRB}
\mathbb{E}\{({\boldsymbol \theta} -\hat{\boldsymbol \theta})({\boldsymbol \theta} -\hat{\boldsymbol \theta})^T\} \geq {\bf C}({\boldsymbol \theta}) = {\bf F}^{-1}({\boldsymbol \theta}),
\end{equation*}
where the Fisher information matrix (FIM) is given by
\begin{equation*}
\label{eq:fim1}
{\bf F}({\boldsymbol \theta}) =  \mathbb{E} \left\{\left(\frac{\partial \ln p({\bf y}; {\boldsymbol \theta})}{\partial {\boldsymbol \theta}}\right) \left(\frac{ \partial\ln p({\bf y}; {\boldsymbol \theta})}{\partial {\boldsymbol \theta}}\right)^T\right\} \in \mathbb{R}^{N \times N},
\end{equation*}
and ${\bf C}({\boldsymbol \theta})$ is the CRB matrix.
An important property of the Fisher information is that it is additive for independent observations, which follows from the fact that
\begin{equation}
\label{eq:property1}
\ln p({\bf y}; {\boldsymbol \theta}) = \ln \prod_{m=1}^M p(y_m; {\boldsymbol \theta}) = \sum_{m=1}^M \ln p(y_m; {\boldsymbol \theta}),
\end{equation}
where we assume that condition (a2) holds.
Using \eqref{eq:property1}, the FIM ${\bf F}({\boldsymbol \theta})$ 
can be alternatively expressed as
\begin{equation*}
\begin{aligned}
\label{eq:fimsum_no}
{\bf F}({\boldsymbol \theta})&= \sum_{m=1}^M \mathbb{E} \left\{\left(\frac{\partial \ln p(y_m; {\boldsymbol \theta})}{\partial {\boldsymbol \theta}}\right) \left(\frac{\partial\ln p(y_m; {\boldsymbol \theta})}{\partial {\boldsymbol \theta}}\right)^T\right\} \end{aligned}
\end{equation*} which can be further simplified to
\begin{equation}
\label{eq:fimsum}
{\bf F}({\boldsymbol \theta}) =  \sum_{m=1}^M {\bf F}_m({\boldsymbol \theta}),
\end{equation} 
where 
\begin{equation}
\label{eq:mFIM}
{\bf F}_m({\boldsymbol \theta}) = 
 \mathbb{E}\left\{ \left(\frac{\partial \ln p(y_m; {\boldsymbol \theta})}{\partial {\boldsymbol \theta}}\right) \left(\frac{\partial\ln p(y_m; {\boldsymbol \theta})}{\partial {\boldsymbol \theta}}\right)^T \right\}
\end{equation}
is the $N \times N$ FIM of the $m$th measurement. In other words, \eqref{eq:fimsum} means that every independent measurement contributes to the information measure. Note that the FIM for non-linear models depends on the unknown vector ${\boldsymbol \theta}$. 

Assume for instance that the observations belong to the family of exponential distributions. The log-likelihood of the observations can then be expressed in the form 
\begin{equation}
\label{eq:exponentialfamily}
\ln \, p({y}_m; {\boldsymbol \theta}) = \ln r({y}_m) + a_m({\boldsymbol \theta})b({y}_m) - c({\boldsymbol \theta}), 
\end{equation}
where $r({y}_m)$ and $b({y}_m)$ are known functions of the observations only, while $a_m({\boldsymbol \theta})$ and $c(\boldsymbol \theta)$ depend only on the unknown parameter. The regularity conditions in general hold for observations that belong to the family of exponential pdfs, and it already includes a large number of distributions.

One specific example that often occurs in practice is the case where the observations $y_m, m =1,2,\ldots,M,$ are related through the following additive Gaussian non-linear model
\begin{equation}
\label{eq:measurementmodel}
{y}_m = {h}_m({\boldsymbol \theta}) + {n}_m, \, m=1,2,\ldots,M, 
\end{equation}
where ${h}_m(\cdot)$ is a non-linear function, and ${n}_m$ is a zero-mean Gaussian random variable with variance $\sigma_m^2$. The log-likelihood of $y_m$ is then given by \eqref{eq:exponentialfamily} with 
\begin{equation*}
\begin{aligned}
r({y}_m) &= \frac{1}{\sqrt{2\pi\sigma_m^2}} \exp(-\frac{1}{2\sigma_m^2} y_m^2),\\
b(y_m) &= y_m/\sigma_m^2,\\
a_m({\boldsymbol \theta}) &= h_m({\boldsymbol \theta}),\\
\text{and} \quad c({\boldsymbol \theta}) &= \frac{1}{2\sigma_m^2} h_m^2({\boldsymbol \theta}).
\end{aligned}
\end{equation*}

Assuming (a2) holds, it is then easy to verify that \eqref{eq:mFIM} simplifies to
$$
{\bf F}_m ({\boldsymbol \theta})=  \frac{1}{\sigma^2_m} \left(\frac{\partial h_m({\boldsymbol \theta})}{\partial {\boldsymbol \theta}}\right) \left(\frac{\partial h_m({\boldsymbol \theta})}{\partial {\boldsymbol \theta}}\right)^T.
$$

\begin{myrem}[Additive Gaussian linear model]
As a special case, when the measurement process is linear, we have 
$y_m = {\bf h}_m^T {\boldsymbol \theta} + n_m, m=1,2,\ldots,M$, i.e., $h_m({\boldsymbol \theta},n_m) := {\bf h}_m^T {\boldsymbol \theta} + n_m$ with $ {\bf h}_m \in \mathbb{R}^N$ being the regressor. The computation of the FIM for a linear model is straightforward, and is given by 
$${\bf F} =  \sum_{m=1}^M \frac{1}{\sigma^{2}_m} {\bf h}_m{\bf h}_m^T.$$ 
The CRB for linear models in additive Gaussian noise is also the MSE, and more importantly it is {\it independent} of the unknown vector.
\end{myrem}
\section{Problem formulation}\label{sec:probfor}
Our goal is now to select the best subset ($\geq N$) of the available $M$ sensor measurements such that a certain accuracy on the estimate $\hat{\boldsymbol \theta}$ is guaranteed.  We next mathematically formulate this sensor selection problem.
\subsection{Sensor selection}\label{sec:passactive}
In order to select the sensors, we introduce a selection vector $${\bf w} = [w_{1},w_2,\ldots,w_{M}]^T \in \{0,1\}^M,$$
where $w_{m} = 1(0)$ indicates that the $m$th sensor measurement is (not) selected. The measurement model including the {\it virtual} hard selection parameter can be visualized as
\begin{equation}
\label{eq:measurementmodelcase1}
{y}_m = {w_{m}}{h}_m({\boldsymbol \theta},{n}_m),  \quad m=1,2,\ldots,M.
\end{equation}
where the selection vector modifies the log-likelihood of the measurements as 
$
\ln \prod_{m=1}^M p(y_m; {\boldsymbol \theta})^{w_m} = \sum_{m=1}^M w_{m} \ln p(y_m; {\boldsymbol \theta}).
$ 
The corresponding FIM matrix in \eqref{eq:fimsum} can then be expressed as
\begin{equation}
\label{eq:fimsel}
\begin{aligned}
{\bf F}({\bf w}, {\boldsymbol \theta}) &=  \sum_{m=1}^M w_{m} {\bf F}_m({\boldsymbol \theta}).
\end{aligned}
\end{equation} 

\subsection{Performance measures}

We  do not restrict ourselves to any specific estimator, however, we use the CRB as a performance measure. The motivation behind using the CRB is as follows:
\begin{itemize}
\item[1.] The CRB is a measure for the (local) identifiability of the problem~\cite{rothenberg1971identification}. More specifically, a non-singular FIM implies (local) solvability and a unique estimate of~${\boldsymbol \theta}$, however, the converse is not necessarily true.  The sensor selection problem presented in this paper seeks a subset of sensors for which the FIM has full rank in some domain such that the solvability of the problem in that domain is always ensured. 
\item[2.] Typically, the subset of selected sensors that yields a lower CRB also yields a lower MSE, and thus improves the performance of any practical system. 
\end{itemize}
The CRB also has a very attractive mathematical structure resulting in a selection problem that can be efficiently solved using convex optimization techniques.

We next use the consistency assumption of the estimator to derive thresholds for the performance measures. We constrain the estimation error ${\boldsymbol \varepsilon } = \hat{\boldsymbol {\theta}} -{\boldsymbol {\theta}}$ to be within an origin-centered circle of radius $R_e$ with a probability higher than $P_e$, i.e., 
\begin{equation}
\label{eq:accuracyrequired}
\mathrm{Pr}({\|\boldsymbol \varepsilon \|}_2 \leq R_e) \geq P_e,
\end{equation}
 where $\mathrm{Pr}(\cdot)$ denotes probability, and the values of $R_e$ and $P_e$ define the accuracy required and are assumed to be known. A higher accuracy level is obtained by reducing $R_e$ and/or  increasing $P_e$. This metric is used in several occasions as an accuracy measure (e.g., see \cite{cover2012elements,Gusta05SPM,Tao09TSP}). We next discuss two popular performance measures that satisfy the above requirement.
 
\subsubsection{Trace constraint}
A sufficient condition to satisfy the accuracy requirement in \eqref{eq:accuracyrequired} is (see Appendix~\ref{app:RSPconstraints})
$$
{\rm tr}\{{\bf C}({\bf w}, {\boldsymbol \theta})\} = {\rm tr}\{ (\sum_{i=1}^M w_m {\bf F}_m(\boldsymbol \theta))^{-1}\} \leq \lambda_{\rm tr}  = (1-P_e)R_e^2.
$$
This measure is related to the {\it A-optimality}. 

\subsubsection{Minimum eigenvalue constraint}
Another popular sufficient condition that also satisfies the accuracy requirement in \eqref{eq:accuracyrequired} is
$$\lambda_{\rm min}\{{\bf F}({\bf w}, {\boldsymbol \theta})\} \geq \lambda_{\rm eig}  = \frac{N}{R_e^2} \left(\frac{1}{1-P_e}\right),$$
where $\lambda_{\rm eig}$ is derived in~\cite{Tao09TSP} (see also Appendix~\ref{app:RSPconstraints}). This measure is related to the {\it E-optimality}. The inequality constraint $\lambda_{\rm min}\{ {\bf F}\} \geq \lambda_{\rm eig}$ can be equivalently expressed as the following linear matrix inequality (LMI): 
\begin{equation}
\label{eq:LMIconstraint}
\sum_{m=1}^{M} w_m {\bf F}_{m}({\boldsymbol \theta})- \lambda_{\rm eig} {\bf I}_N  \succeq {\bf 0}_N.
\end{equation}
In other words, we put a lower bound on each eigenvalue of the matrix ${\bf F}$. The solution set of ${\bf w}$ satisfying this LMI is convex as ${\bf F}_m({\boldsymbol \theta}) \in \mathbb{S}^N, m=1,2,\ldots,M$ and $\lambda_{\rm eig} {\bf I}_N \in \mathbb{S}^N$~\cite[Pg. 38]{Boyd}.  

The trace constraint has a larger feasible set as compared to the minimum eigenvalue constraint. However, although the trace constraint is a sufficient condition, the resulting sensor selection problem is computationally less attractive compared to the minimum eigenvalue constraint (as we show later). Moreover, LMIs can be used to also represent the trace constraint. For these reasons, we focus on the minimum eigenvalue (LMI) constraints from now on. However, without loss of generality (w.l.o.g.) either one of the two performance constraints can be used.    

The above performance measures depend on the unknown parameter. In practice, the unknown parameter ${\boldsymbol \theta}$ has a physical meaning and takes values within a certain domain denoted by $\mathcal{U}$. For example, in the case of direction-of-arrival estimation, $\mathcal{U}$ is the sector where the source is expected or for target localization it is the surveillance area where the target resides. Since the FIM for non-linear models depends on the unknown ${\boldsymbol \theta}$, we propose to constrain every point within the domain $\mathcal{U}$. 
\begin{myrem}[Bayesian CRB constraint]
In a Bayesian setting, when {\it prior} information of the unknown parameter ${\boldsymbol \theta}$ is available,  this additional knowledge typically yields a lower CRB, and the related information matrix is often called the Bayesian information matrix (BIM). The BIM is given by ${\bf F}_{\rm B} ({\bf w}, {\boldsymbol \theta})= {\bf F}({\bf w}, {\boldsymbol \theta})+ {\bf J}_{\rm p}$, where ${\bf J}_{\rm p}$ is some prior information matrix ${\bf J}_{\rm p} =- \mathbb{E}_{\boldsymbol \theta} \left\{\frac{\partial}{\partial{\boldsymbol \theta}} \left(\frac{\ln p({\boldsymbol \theta})}{\partial {\boldsymbol \theta}}\right)^T\right\}$ with the (log) {\it prior} $\ln p({\boldsymbol \theta})$. The LMI constraint in \eqref{eq:LMIconstraint} for the Bayesian setting will then be 
\begin{equation}
\label{eq:Blmi}
{\bf J}_{\rm p} + \sum_{m=1}^{M} w_m {\bf F}_{m}({\boldsymbol \theta})  \succeq \lambda_{\rm eig} {\bf I}_N.
\end{equation}
The prior information typically comes from the dynamics, previous measurements, or combining other available measurements.
\end{myrem}

\subsection{Problem statement}
Having introduced the selection vector as well as the performance measure we can now formally state the problem.
\begin{myproblem}[Sensor selection]
Given the likelihoods $p(y_m; {\boldsymbol \theta}), m=1,2,\ldots,M,$ of the measurements, and assuming (a1) and (a2) hold, find a vector ${\bf w} \in \{0,1\}^M$ that selects the minimum number of most informative sensors satisfying the performance measure $\sum_{m=1}^{M} w_m {\bf F}_{m}({\boldsymbol \theta}) - \lambda_{\rm eig} {\bf I}_N\succeq {\bf 0}_N, \,\; \forall {\boldsymbol \theta} \, \in \, \mathcal{U}$. 
\end{myproblem}

In order to reduce the hardware costs, storage, processing, and communication overhead, we minimize the number of selected sensors. This can be achieved by minimizing the cardinality of the selection vector, i.e., by minimizing the number of non-zero entries of the selection vector.  Mathematically, the {\it sensor selection problem} is  formulated as the design of a selection vector which can be expressed as the following optimization problem
\begin{subequations}
 \label{eq:l0}
\begin{align} 
{\bf w}^\ast =   \hskip0.5mm &\argmin_{\bf w}  \quad {\|{\bf w}\|}_0\label{eq:l0a}\\
&\hskip1mm{\rm s.t.}  \quad \sum_{m=1}^{M} w_m {\bf F}_{m}({\boldsymbol \theta}) - \lambda_{\rm eig} {\bf I}_N \succeq {\bf 0}_N, \quad\forall {\boldsymbol \theta} \, \in \, \mathcal{U} \label{eq:l0b},\\
&\hskip10mm {{\bf w}\, \in \, \{0,1\}^M},\label{eq:l0c}
\end{align}
\end{subequations}
where the $\ell_0$-(quasi) norm refers to the number of non-zero entries in ${\bf w}$, i.e., ${\|{\bf w}\|}_0 := |\{m\,: \, w_m \neq 0\}|$. The threshold $\lambda_{\rm eig}$ imposes the accuracy requirement. The threshold $\lambda_{\rm eig}$ is also the sparsity-inducing parameter, where $\lambda_{\rm eig}\rightarrow 0$ implies a sparser solution.  

Suppose the domain $\mathcal{U}$ consists of $D$ points, obtained by gridding the entire domain at a certain resolution. The resulting multiple LMI constraints can be stacked together as a single LMI constraint. Let us consider the domain $\mathcal{U} = \{{\boldsymbol \theta}_1,{\boldsymbol \theta}_2,\ldots,{\boldsymbol \theta}_D\}$ with $|\mathcal{U}| = D$. The constraints in \eqref{eq:l0b} can then be equivalently expressed as a single LMI constraint written as
$\sum_{m=1}^{M} w_m {\bf F}_{m} - \lambda_{\rm eig} {\bf I}_{DN} \succeq {\bf 0}_{DN}$, where ${\bf F}_m = {\rm diag}({\bf F}_m({\boldsymbol \theta}_1),{\bf F}_m({\boldsymbol \theta}_2),\ldots,{\bf F}_m({\boldsymbol \theta}_D)) \in \mathbb{S}^{DN}$ for $m=1,2,\ldots,M$. Note that the FIM after gridding is independent of ${\boldsymbol \theta}$, and we denote this simply by ${\bf F}_m$ (not explicitly as a function of ${\boldsymbol \theta}$). 
 
\begin{myrem} [Worst-case constraints]
If there exists some $\tilde{\boldsymbol{\theta}} \in \mathcal{U}_w \subset \mathcal{U}$ such that $\lambda_{\rm min}({\bf F}({\bf w},\tilde{\boldsymbol \theta})) \leq \lambda_{\rm min}({\bf F}({\bf w},{\boldsymbol \theta})),\, \forall{\bf w} \in \{0,1\}^M$ and $\forall {\boldsymbol \theta} \in \mathcal{U}_w$, then it is sufficient to constrain the performance for only the worst-case $\tilde{\boldsymbol \theta} \in \mathcal{U}_w$ instead of $\forall {\boldsymbol \theta} \in \mathcal{U}_w$. This property can be used a guideline for gridding.
\end{myrem}
\section{Sensor selection solvers}\label{sec:optprob}
It is well known that the $\ell_0$-(quasi) norm optimization is NP-hard and nonconvex. More specifically, the original sensor selection problem in \eqref{eq:l0} is {NP-hard}. The Boolean constraint in \eqref{eq:l0c} is non-convex and incurs a combinatorial complexity. We next present a number of solvers with which the relaxed convex problem can be solved efficiently in polynomial time.     

\subsection{Convex approximation based on $\ell_1$-norm}
A computationally tractable (suboptimal) solution is to use the traditional best convex surrogate for the $\ell_0$-(quasi) norm namely the $\ell_1$-norm heuristic. The $\ell_1$-norm is known to represent an efficient heuristic for the $\ell_0$-(quasi) norm optimization with convex constraints especially when the solution is sparse~\cite{Polyakl1convex2013}. Such relaxations are well-studied for problems with linear constraints in the context of compressed sensing (CS) and sparse signal recovery~\cite{donoho2006most}. The non-convex Boolean constraint in \eqref{eq:l0c} is further relaxed to the convex box constraint $[0,1]^M$. 

The {\it relaxed} optimization problem is given as the following SDP problem
\begin{subequations}
\label{eq:l1}
\begin{align} 
\hat{\bf w} =\hskip0.5mm &\argmin_{{\bf w}\, \in \, \mathbb{R}^M} \quad {\|{\bf w}\|}_1\label{eq:l1a}\\
&\hskip1mm{\rm s.t.} \, \sum_{m=1}^{M} w_m {\bf F}_{m} - \lambda_{\rm eig} {\bf I}_{DN} \succeq {\bf 0}_{DN}, \label{eq:l1b}\\ 
&\hskip8mm {0} \leq {w}_m \leq {1}, \quad m=1,2,\ldots,M, \label{eq:l1c}
\end{align}
\end{subequations}
where ${\|{\bf w}\|}_1 = \sum_{m=1}^M |w_m|$ denotes the $\ell_1$-norm. 
Due to the positivity constraint, the objective function $\|{\bf w}\|_1$ will simply be an affine function ${\bf 1}_M^T{\bf w}$. 
The optimization problem in \eqref{eq:l1} is a standard SDP problem in the inequality form, which can be efficiently solved in polynomial time using interior-point methods~\cite{Boyd}. 
An implementation of the interior-point method for solving SDP problems in the inequality form is typically based on Newton's method using an approximating barrier function. A brief description of the {\it projected Newton's method} is provided in Appendix~\ref{app:Newton's} which is used to analyze the computational complexity of the relaxed sensor selection problem.
\begin{myrem}[Complexity per iteration]
The computational cost involved during each iteration is as follows~\cite[Pg. 619]{Boyd}. The matrices ${\bf F}_m, m =1,2,\ldots,M,$ have a block-diagonal structure with $D$ blocks. Forming the matrix ${\bf S} = \sum_{m=1}^{M} w_m {\bf F}_{m} - \lambda_{\rm eig} {\bf I}_{DN}$ costs $O(DMN^2)$ flops; computing ${\bf S}^{-1}{\bf F}_i \; \forall i$ via Cholesky factorization costs $O(MDN^3)$ flops; the Hessian matrix is computed via the inner product of the matrices ${\bf S}^{-1}{\bf F}_i$ and ${\bf S}^{-1}{\bf F}_j$, which costs $O(DM^2N^2) \; \forall i,j$. Finally, the Newton step is computed via Cholesky factorization costing $O(M^3)$ flops, and the projection costs $O(M)$ flops. Assuming that $M \gg N$, the overall computational complexity per iteration of the projected Newton's algorithm is then $O(M^3)$.
\end{myrem}
 Implementations of the interior-point methods are easily available in the form of well-known toolboxes like Yalmip~\cite{YALMIP}, SeDuMi~\cite{Sturm98usingsedumi}, and CVX~\cite{cvx}.

\subsection{Projected subgradient algorithm} \label{sec:projsub}
The second-order Newton's method (cf. Appendix~\ref{app:Newton's}) is typically intractable when the number of sensors is very large ($M \gg 1000$ for example). To circumvent this problem, we propose a subgradient based algorithm. The projected subgradient algorithm is a first-order method which is attractive for large-scale problems as each iteration is much cheaper to process. 

The subgradient method is typically used for optimizations involving non-differentiable functions~\cite{boyd2003subgradient,bertsekas1999nonlinear}. The subgradient method is a generalization of the gradient method for non-smooth and non-differentiable functions, such as, the $\ell_1$-norm and the minimum eigenvalue constraint functions. We next derive the projected subgradient algorithm.

The relaxed sensor selection problem in \eqref{eq:l1} can be equivalently expressed as
\begin{subequations}
\label{eq:psg1}
\begin{align} 
&\argmin_{{\bf w}} \quad {\|{\bf w}\|}_1\label{eq:psga}\\
&\hskip1mm{\rm s.t.} \, f_{\rm eig}({\bf w}) \geq \lambda_{\rm eig}, \label{eq:psgb}\\ 
&\hskip8mm {\bf w} \in \mathcal{W}, \label{eq:psgc}
\end{align}
\end{subequations}
where $f_{\rm eig}({\bf w}) := \lambda_{\rm min}\{\sum_{m=1}^{M} w_m {\bf F}_{m}\}$ is the constraint function in \eqref{eq:l1b}, and the set $\mathcal{W} = \{{\bf w} \in \mathbb{R}^M \mid 0 \leq w_m \leq 1, m=1,2,\ldots,M\}$ denotes the box constraints in \eqref{eq:l1c}.

The objective ${\bf 1}_M^T{\bf w}$ is affine, so a subgradient of the objective is the all-one vector ${\bf 1}_M$. Let ${\bf g}^{k} \in \partial f_{\rm eig}({\bf w}^{k})$ denote a subgradient of the constraint function $f_{\rm eig}({\bf w})$ at ${\bf w} = {\bf w}^{k}$. Here, the set $\partial f_{\rm eig}({\bf w}^{k})$ denotes the subdifferential of $f_{\rm eig}({\bf w})$ evaluated at ${\bf w} = {\bf w}^{k}$. To compute ${\bf g}^{k}$, we express the constraint function $f_{\rm eig}({\bf w}^{k})$ as
$$
f_{\rm eig}({\bf w}^{k}) = \inf_{\| {\bf v}\|\leq 1} \quad {\bf v}^T\left(\sum_{m=1}^{M} w_m^{k} {\bf F}_{m}\right){\bf v}.
$$
The computation of a subgradient is straightforward, and is given by
$$ 
{\bf g}^{k} = [({\bf v}^{k}_{\rm min})^T{\bf F}_1{\bf v}^k_{\rm min},\ldots,({\bf v}^k_{\rm min})^T{\bf F}_m{\bf v}^k_{\rm min}]^T \in \partial f_{\rm eig}({\bf w}^{k}),
$$
where ${\bf v}^k_{\rm min}$ is the eigenvector corresponding to the minimum eigenvalue $\lambda_{\rm min}\{\sum_{m=1}^{M} w_m^k {\bf F}_{m}\}$. The minimum eigenvalue and the corresponding eigenvector can be computed using a low-complexity iterative algorithm called the power method (see Appendix~\ref{app:poweriter}) or using the standard eigenvalue decomposition~\cite{golub1996matrix}. Let the projection of a point onto the set $\mathcal{W}$ be denoted by $\mathcal{P}_{\mathcal{W}}(\cdot)$, which can be expressed elementwise as
\begin{equation}
\label{eq:projection}
[\mathcal{P}_{\mathcal{W}}({\bf w})]_m = \begin{cases} 0 & \mbox{if } w_m \leq 0, \\ 
w_m & \mbox{if } 0 < w_m < 1, \\ 
1 & \mbox{if }w_m \geq 1.  \end{cases}
\end{equation}
The {\it projected subgradient algorithm} then proceeds as follows:
\begin{equation}
\small
\label{eq:subprjupdate}
\begin{aligned}
{\bf w}^{k+1} = \begin{cases} \mathcal{P}_{\mathcal{W}}({\bf w}^{k} - \alpha^{k}{\bf 1}_M) & \mbox{if } f_{\rm eig}({\bf w}^{k}) \geq  \lambda_{\rm eig}, \\ 
\mathcal{P}_{\mathcal{W}}({\bf w}^{k} +\alpha^{k}{\bf g}^{k}) & \mbox{if }f_{\rm eig}({\bf w}^{k}) <  \lambda_{\rm eig}.  \end{cases}
\end{aligned}
\end{equation}
In other words, if the current iterate ${\bf w}^{k}$ is feasible (i.e., $f_{\rm eig}({\bf w}^{k}) \geq  \lambda_{\rm eig}$), we update ${\bf w}$ in the direction of a negative objective subgradient, as if the LMI constraints were absent; If the current iterate ${\bf w}^{k}$ is infeasible (i.e., $f_{\rm eig}({\bf w}^{k}) <  \lambda_{\rm eig}$), we update ${\bf w}$ in the direction of a subgradient ${\bf g}^{k}$ associated with the LMI constraints. After the update is computed, the iterate is projected onto the constraint set $\mathcal{W}$ using $\mathcal{P}_{\mathcal{W}}(\cdot)$. 

\begin{algorithm}[!t]
\caption{Projected subgradient algorithm}
\label{alg:projsubalgo}
\begin{algorithmic}
\item[1.] \textbf{Initialize} iteration counter $k=0$, ${\bf w}^{k} = {\bf 1}_M$, ${\bf g}^{k} = {\bf 0}$, ${k}_{\rm max}$, $\epsilon$, and $\lambda_{\rm eig}$.
\item[2.] \textbf{for} $k=0$ to $k_{\rm max}$
\item[3.] \hskip1cm\textbf{compute} $f_{\rm eig}({\bf w}^{k}) = \lambda_{\rm min}\{\sum_{m=1}^{M} w_m^k {\bf F}_{m}\}$
\item[4.] \hskip1cm\textbf{update}
\item[5.] \hskip1.5cm\textbf{if} $f_{\rm eig}({\bf w}^{k}) \geq \lambda_{\rm eig}$
\item[6.] \hskip2cm ${\bf w}^{k+1}=\mathcal{P}_{\mathcal{W}}({\bf w}^{k} - (1/\sqrt{k}){\bf 1}_M)$
\item[7.] \hskip1.5cm\textbf{elseif} $f_{\rm eig}({\bf w}^{k}) < \lambda_{\rm eig}$
\item[8.]  \hskip2cm ${\bf w}^{k+1}=\mathcal{P}_{\mathcal{W}}({\bf w}^{k} +\frac{f_{\rm eig}({\bf w}^{k}) + \epsilon}{\|{\bf g}^{k}\|_2^2}{\bf g}^{k})$
\item[9.] \hskip1.5cm\textbf{end}
\item[10.] \textbf{end} 
\item[11.] $\hat{\bf w} = {\bf w}^{{k}_{\rm max}}$
\end{algorithmic}
\end{algorithm} 
When the $k$th iterate is feasible, a diminishing non-summable step size $\alpha^{k} = 1/\sqrt{k}$ is used. When the iterate is not feasible Polyak's step size $\alpha^{k} = \frac{f_{\rm eig}({\bf w}^{k}) + \epsilon}{\|{\bf g}^{k}\|_2^2}$ is used, where we adopt the optimal value for $\epsilon := {\bf 1}_M^T{\bf w}^\ast$ when $\|{\bf w}\|_0$ known (i.e., the number of sensors to be selected is known). If this is not known, then we approximate it with $\epsilon := f_{\rm best}^{k} + \gamma$, where  $\gamma = 10/ (10+k)$, and $f_{\rm best}^{k} = \min\{f_{\rm best}^{k-1}, {\bf 1}_M^T{\bf w}^{k}\}$~\cite{boyd2003subgradient}.
The algorithm is terminated after a specified maximum number of iterations $k_{\rm max}$. Finally, the estimate is denoted by $\hat{\bf w} = {\bf w}^{k_{\rm max}}$.

The convergence results of the subgradient method for the constrained optimization (i.e., without the projection step) are derived in~\cite{boyd2003subgradient}. Since the projection onto a convex set is non-expansive~\cite{bertsekas1999nonlinear}, it does not affect the convergence. 
The projected subgradient algorithm is summarized as Algorithm~\ref{alg:projsubalgo}.
 
\begin{myrem}[Complexity per iteration]
We first form the matrix $\sum_{m=1}^{M} w_m {\bf F}_{m}$, which costs $O(DMN^2)$ flops. 
The minimum eigenvalue and the corresponding eigenvector can be computed using the power method at a cost of $O(DN^2)$ flops~\cite{golub1996matrix}. Forming the vector ${\bf g}$ costs $O(DMN^2)$ flops, computing its norm costs $O(M)$ flops, and the update and projection together cost $O(M)$ flops. Assuming that $M \gg N$ as earlier, the  computational cost of the projected subgradient algorithm is $O(DMN^2)$ which is much lower than the complexity of the projected Newton's method.
\end{myrem}

A {\it distributed implementation} of the projected subgradient algorithm is very easy. A simple distributed averaging algorithm (e.g., \cite{xiao2004fast}) can be used to compute the sum of matrices $\sum_{m=1}^{M} w_m {\bf F}_{m}$. The minimum eigenvalue and the corresponding eigenvector can then be computed using power iterations at each node independently. The update equation \eqref{eq:subprjupdate}, the subgradient vector ${\bf g}$, and the projection are computed coordinatewise and are {\it already} distributed. 

Subgradient methods are typically very slow compared to the interior-point method involving Newton iterations, and subgradient methods typically require a few hundred iterations. Newton's method typically requires in the order of ten steps. On the other hand, unlike the projected subgradient method, Newton's method cannot be easily distributed, and requires a relatively high complexity per iteration due to the computation and storage of up to second-order derivatives. Depending on the scale of the problem and the resources available for processing one could choose between the subgradient or Newton's algorithm. 

\subsection{Concave surrogate: sparsity-enhancing iterative algorithm}\label{sec:iter1}
The $\ell_1$-norm is customarily used as the best convex relaxation for the $\ell_0$-norm. However, the intersection of the $\ell_1$-norm ball (or an affine subspace) with the positive semi-definite cone (i.e., the LMI constraint) is not always a unique point as shown in the following Theorem.   

\begin{mytheo}[Uniqueness]\label{eq:lemunique}
The projection of a point ${\bf w} \in [0,1]^M$ onto a convex LMI constraint set $\sum_{m=1}^{M} w_m {\bf F}_{m} - \lambda_{\rm eig} {\bf I}_{DN} \succeq {\bf 0}_{DN}$ under the $\ell_1$-norm is not always unique. 
\end{mytheo}

\begin{IEEEproof}
The proof follows from the fact that the $\ell_1$-norm is not strictly convex, and from the linearity of the constraint set.  Let us consider an example with $M=2$ (w.l.o.g.), and ${\bf F}_1 = {\bf F}_2 \succeq \lambda_{\rm eig}{\bf I}_{DN}$. In other words, the observations are {\it identical}. In this case, the extreme points of the $\ell_1$-norm ball, i.e., $\hat{\bf w}_1 = (1,0)$ and $\hat{\bf w}_2 = (0,1)$ are two example solutions. Moreover, since the solution set of a convex minimization problem is convex, $\tau \hat{\bf w}_1 + (1-\tau)\hat{\bf w}_2$ is also a solution for {\it any} $0 < \tau <  1$, which gives an infinite number of solutions to the relaxed optimization problem \eqref{eq:l1}. For such cases, the $\ell_1$-norm relaxation will typically not result in a sparse solution.
\end{IEEEproof}
To improve upon the $\ell_1$-norm solution due to its non-uniqueness following from Theorem~\ref{eq:lemunique}, we propose an alternative relaxation for the original sensor selection problem which also results in fewer selected sensors. Instead of relaxing the $\ell_0$-(quasi) norm with the $\ell_1$-norm, using a nonconvex surrogate function can yield a {\it better} approximation.  It is motivated in~\cite{candes08} that the logarithm of the geometric mean of its elements can be used as an alternative surrogate function for linear inverse problems in CS. Adapting this to our sensor selection problem, we arrive at the optimization problem
\begin{subequations}
\label{eq:concavesurr}
\begin{align} 
\hskip0.5mm &\argmin_{{\bf w}\, \in \, \mathbb{R}^M} \quad \sum_{m=1}^M \ln \,(w_m + \delta)\label{eq:concavesurra}\\
&\hskip1mm{\rm s.t.} \, \sum_{m=1}^{M} w_m {\bf F}_{m} - \lambda_{\rm eig} {\bf I}_{DN} \succeq {\bf 0}_{DN},\\
&\hskip8mm {0} \leq {w}_m \leq {1}, \quad m=1,2,\ldots,M, \label{eq:concavesurrc}
\end{align}
\end{subequations}
where $\delta > 0$ is a small constant that prevents the cost from tending to $- \infty$. The cost \eqref{eq:concavesurra} is concave, but since it is smooth w.r.t. ${\bf w}$, iterative linearization can be performed to obtain a local minimum~\cite{candes08}. 
The first-order approximation of $\ln \,(w_m + \delta)$ around $(w_m[i-1] + \delta)$ results in 
$$
\ln \,(w_m + \delta) \leq \ln \,(w_m[i-1] + \delta) + \frac{(w_m-w_m[i-1])}{(w_m[i-1] + \delta)}.
$$
Instead of minimizing the original cost, the majorizing cost (second term on the right-hand side of the above inequality) can be optimized to attain a local minima. More specifically, the optimization problem \eqref{eq:concavesurr} can be iteratively driven to a local minimum using the iterations
\begin{subequations}
\label{eq:convexiter}
\begin{align} 
\hat{\bf w}[i]=\hskip0.5mm &\argmin_{{\bf w}\, \in \, \mathbb{R}^M} \quad \sum_{m=1}^M \frac{w_m}{\hat{w}_{m}[i-1]+\delta}\label{eq:convexitera}\\
&\hskip1mm{\rm s.t.} \, \sum_{m=1}^{M} w_m {\bf F}_{m} - \lambda_{\rm eig} {\bf I}_{DN} \succeq {\bf 0}_{DN}, \label{eq:convexiterb}\\
&\hskip8mm {0} \leq {w}_m \leq {1}, \quad m=1,2,\ldots,M. \label{eq:convexiterc}
\end{align}
\end{subequations}
The iterative algorithm is summarized as Algorithm~\ref{alg:iterativealgo}. Each iteration in \eqref{eq:convexiter} solves a weighted $\ell_1$-norm optimization problem. The weight updates force the small entries of the vector $\hat{\bf w}[i]$ to zero and avoid inappropriate suppression of larger entries. The parameter $\delta$ provides stability, and guarantees that the zero-valued entries of $\hat{\bf w}[i]$ do not strictly prohibit a nonzero estimate at the next step. Finally, the estimate is given by $\hat{\bf w} = \hat{\bf w}[i_{\rm max}]$, where $i_{\rm max}$ is the specified maximum number of iterations.
\begin{myrem}[Sparsity-enhancing projected subgradient algorithm]
The projected subgradient algorithm can be adapted to fit into the sparsity-enhancing iterative algorithm as well. The optimization problem \eqref{eq:l1iter} is then replaced with the following update equations:
\begin{equation*}
\small
\label{eq:subprjupdatesparse}
\begin{aligned}
{\bf w}^{k+1}[i] = \begin{cases} \mathcal{P}_{\mathcal{W}}({\bf w}^{k}[i] - \alpha^k{\bf u}[i]) & \mbox{if } f_{\rm eig}({\bf w}^{k}[i]) \geq \lambda_{\rm eig}, \\ 
\mathcal{P}_{\mathcal{W}}({\bf w}^{k}[i] + \alpha^k{\bf g}^{k}[i]) & \mbox{if }f_{\rm eig}({\bf w}^{k}[i]) < \lambda_{\rm eig},  \end{cases}
\end{aligned}
\end{equation*} 
where we solve a number of iterations (inner loop) of the projected subgradient algorithm within the $i$th iteration (outer loop) of Algorithm~\ref{alg:iterativealgo}. Here, the $k$th iterate of the inner loop in the $i$th outer loop is denoted as $(\cdot)^k[i]$.
\end{myrem}

From the solution of the relaxed optimization problem, the approximate Boolean solution to ${\bf w} \in \{0,1\}^M$ can be obtained using randomization techniques, as described next. 
\begin{algorithm}[!t]
\caption{Sparsity-enhancing iterative algorithm}
\label{alg:iterativealgo}
\begin{algorithmic}
\item[1.] \textbf{Initialize} the iteration counter $i=0$, the weight vector ${\bf u}[0]= [u_1[0],u_2[0],\ldots,u_M[0]]^T = {\bf 1}_M$, $\delta$, and $i_{\rm max}$.
\item[2.] \textbf{for} $i=0$ to $i_{\rm max}$
\item[3.] \hskip1cm\textbf{solve} the weighted $\ell_1$-norm minimization problem
\begin{subequations}
\label{eq:l1iter}
\begin{align} 
\hat{\bf w}[i] =\hskip0.5mm &\argmin_{{\bf w}\, \in \, \mathbb{R}^M} \quad {\bf u}[i]^{T}{\bf w} \label{eq:l1itera}\\
&\hskip1mm{\rm s.t.} \, \sum_{m=1}^{M} w_m {\bf F}_{m} - \lambda_{\rm eig} {\bf I}_{DN} \succeq {\bf 0}_{DN}, \label{eq:l1iterb}\\ 
&\hskip8mm {0} \leq {w}_m \leq {1}, m=1,2,\ldots,M.\label{eq:l1iterc}
\end{align}
\end{subequations}
\item[4.] \hskip1cm\textbf{update}  the weight vector $u_m[i+1] = \frac{1}{\delta + \hat{w}_{m}[i]}$, for each $m=1,2,\ldots,M$.
\item[5.] \textbf{end}
\item[6.] $\hat{\bf w} = \hat{\bf w}[i_{\rm max}]$.
\end{algorithmic}
\end{algorithm} 
\subsection{Randomized rounding}
\begin{algorithm}[!t]
\caption{Randomized rounding algorithm}
\label{alg:randomizedround}
\begin{algorithmic}
\item[1.]  Generate $l=1,2,\ldots,L,$ candidate estimates of the form $\hat{w}_{m,l} =1$ with a probability 
$\hat{w}_{m}$ (or $\hat{w}_{m,l} =0$ with a probability 
$1- \hat{w}_{m}$) for $m=1,2,\ldots,M$. 
\item[2.] Define $\hat{\bf w}_{l} = [\hat{w}_{1,l},\ldots,\hat{w}_{M,l}]^T$ and the index set of the candidate estimates satisfying the constraints as 
$${\Omega} \triangleq \{l \mid \lambda_{\rm min}\{{\bf F}(\hat{\bf w}_l, {\boldsymbol \theta})\} \geq \lambda_{\rm eig} , \forall {\boldsymbol \theta} \, \in \, \mathcal{U},\, l=1,2,\ldots,L\}.$$
\item[3.] If the set $\Omega$ is empty, go back to step 1.
\item[4.] The suboptimal Boolean estimate is the solution to the optimization problem
$${\hat{\bf w}}_{\rm bp} =\argmin_{l \in \Omega} {\| \hat{\bf w}_{l} \|}_1.$$
\end{algorithmic}
\end{algorithm}
The solution of the relaxed optimization problem is used to compute the suboptimal Boolean solution for the selection problem.
 A straightforward technique that is often used is the simple rounding technique, in which the Boolean estimate is given by ${\rm round}(\hat{w}_{m}), \; m=1,2,\ldots,M,$
where we define $\hat{\bf w} \triangleq [\hat{w}_{1},\hat{w}_{2},\ldots,\hat{w}_{M}]^T$, and the ${\rm round}(.)$ operator rounds its arguments towards the nearest integer. However, there is no guarantee that the Boolean estimates obtained from the rounding technique always satisfy the LMI constraint. Hence, we propose a randomized rounding technique, where the suboptimal Boolean estimates are computed based on random experiments guided by the solution from the SDP problem in \eqref{eq:l1} or the iterative version in \eqref{eq:convexiter}. The randomized rounding technique is summarized as Algorithm~\ref{alg:randomizedround}.
\section{Extensions}\label{sec:extension}
\subsection{The dual problem}\label{sec:analysis}
The dual of the relaxed primal optimization problem has an interesting relation to the diameter of the confidence ellipsoids, and is closely related to the dual of the {\it E-optimal design}~\cite[Pg. 388]{Boyd}. The dual problem of \eqref{eq:l1} is given as follows
\begin{equation}
\label{eq:dual}
\begin{aligned} 
\hskip-1mm(\hat{\bf Z}, \hat{\boldsymbol \mu}) =\hskip0.5mm &\argmax_{{\bf Z}, \, {\boldsymbol \mu}} \quad \lambda_{\rm eig}\,{\rm tr}\{{\bf Z}\}-{\bf 1}_M^T {\boldsymbol \mu}\\
&{\rm s.t.} \quad \mathbb{E} \{{\bf s}_m^T{\bf Z}{\bf s}_m\} \leq 1 + \mu_m,  m=1,2,\ldots,M,\\
&\hskip10mm {\bf Z} \succeq {\bf 0}, \, \mu_m \geq  0, \, m=1,2,\ldots,M, 
\end{aligned}
\end{equation}
where ${\bf Z} \in \mathbb{S}^{DN}_{+}$ and ${\boldsymbol \mu} \in \mathbb{R}^N$ are the dual variables, and we use 
$\mathbb{E} \{{\bf s}_m^T{\bf Z}{\bf s}_m\}={\rm tr}\{{\bf F}_m{\bf Z}\}$ with ${\bf s}_m = [\frac{\partial h_m(\tilde{\boldsymbol \theta}_1,n_m)}{\partial \tilde{\boldsymbol \theta}_1^T},\ldots,\frac{\partial h_m(\tilde{\boldsymbol \theta}_D,n_m)}{\partial \tilde{\boldsymbol \theta}_D^T}]^T \in \mathbb{R}^{DN}$. For a detailed derivation of the dual problem, see Appendix~\ref{app:dual}.
The dual problem can be interpreted as the problem of maximizing the (average) diameter of the {\it confidence ellipsoid}.
If we set $\mu_m=0, m=1,2,\ldots,M$, the optimal solution $\hat{\bf Z}$ to the problem \eqref{eq:dual} is also the solution to the dual of the {\it E-optimal} design problem~\cite[Pg. 388]{Boyd}, which maximizes the diameter of the confidence ellipsoid centered around the origin.
 

The dual formulation is often easier to solve and has only $M$ inequality constraints. The dual problem can be solved using Yalmip, SeDuMi, or CVX as earlier. Suppose ${\bf Z}$ and ${\boldsymbol \mu}$ are dual feasible, and ${\bf w}$ is primal feasible, then the dual problem yields the following bound on the primal SDP problem:
$\lambda_{\rm eig}\,{\rm tr}\{{\bf Z}\}-{\bf 1}_M^T {\boldsymbol \mu} \leq {\bf 1}_M^T{\bf w}.$

\subsection{Scalar constraints}
\subsubsection*{Trace constraint}
The relaxed sensor selection problem with the scalar trace constraint is given as follows
\begin{equation}
\label{eq:l1trace}
\begin{aligned} 
\hskip0.5mm &\argmin_{{\bf w}\, \in \, \mathbb{R}^M} \quad {\|{\bf w}\|}_1\\
&\hskip1mm{\rm s.t.} \, {\rm tr}\{(\sum_{m=1}^{M} w_m {\bf F}_{m}({\boldsymbol \theta}))^{-1}\} \leq \lambda_{\rm tr}, \quad \forall {\boldsymbol \theta} \in \mathcal{U} , \\ 
&\hskip8mm {0} \leq {w}_m \leq {1}, \quad m=1,2,\ldots,M.
\end{aligned}
\end{equation}
The trace constraint in \eqref{eq:l1trace} is convex in ${\bf w}$; this is easier to verify when the above trace constraint is expressed as an LMI~\cite[Pg. 387]{Boyd}. The optimization problem in \eqref{eq:l1trace} is a convex problem, and can be cast as an SDP:
\begin{equation}
\label{eq:l1traceSDP}
\begin{aligned} 
\hskip0.5mm &\argmin_{{\bf w} \in \mathbb{R}^M, \, {\bf x} \in \mathbb{R}^N} \quad {\|{\bf w}\|}_1\\
&\hskip1mm{\rm s.t.} \, \left[\begin{array}{cc}\sum_{m=1}^{M}w_m {\bf F}_{m}({\boldsymbol \theta}) & {\boldsymbol \delta}_n \\{\boldsymbol \delta}_n^T & x_n\end{array}\right] \succeq {\bf 0}_{N+1}, n=1,2,\ldots,N,  \\ 
&\hskip8mm {0} \leq {w}_m \leq {1}, \quad m=1,2,\ldots,M, \\
& \hskip8mm {\bf 1}_N^T{\bf x} \leq \lambda_{\rm tr}, \, x_n \geq 0, n=1,2,\ldots,N, \, \forall {\boldsymbol \theta} \in \mathcal{U},
\end{aligned}
\end{equation}
where ${\bf x} = [x_1,x_2,\ldots,x_N]^T \in \mathbb{R}^N$ is a variable, and ${\boldsymbol \delta}_n$ is the $n$th unit vector in $\mathbb{R}^N$. 
The optimization problem in \eqref{eq:l1traceSDP} has $N$ LMI constraints for every point in $\mathcal{U}$ and $N+1$ inequality constraints (in addition to the box constraint), while the optimization problem in \eqref{eq:l1} has only one LMI constraint for every point in $\mathcal{U}$ (in addition to the box constraint). Hence, solving  \eqref{eq:l1traceSDP} is computationally more intense than solving \eqref{eq:l1}.

\subsubsection*{Determinant constraint}
Another popular scalar performance measure for the quality of the estimate is the determinant (product of eigenvalues) constraint. This measure is related to the D-optimality.
The relaxed sensor selection problem with the determinant constraint is given as follows
\begin{equation}
\label{eq:l1determinant}
\begin{aligned} 
\hskip0.5mm &\argmin_{{\bf w}\, \in \, \mathbb{R}^M} \quad {\|{\bf w}\|}_1\\
&\hskip1mm{\rm s.t.} \, \ln {\rm det}\{\sum_{m=1}^{M} w_m {\bf F}_{m}({\boldsymbol \theta})\} \geq  \lambda_{\rm det}, \forall{\boldsymbol \theta} \in \mathcal{U}, \\ 
&\hskip8mm {0} \leq {w}_m \leq {1}, \, m=1,2,\ldots,M, 
\end{aligned}
\end{equation}
where the threshold $\lambda_{\rm det}$ specifies the mean radius a confidence ellipsoid (see Appendix~\ref{app:RSPconstraints}). In other words, although it is an indication of the performance of the estimator, it is not a sufficient condition for \eqref{eq:accuracyrequired}.
The log-determinant constraint is a concave function of ${\bf w}$ for $w_m \geq 0,$ for $m=1,2,\ldots,M$. 

The relaxed sensor selection problem with the scalar (trace or determinant) constraints can be solved with either one of the two proposed cost functions, i.e., the $\ell_1$-norm or the log-based concave surrogate.

\section{Examples: Sensor placement for localization}\label{sec:example}
Localization is an important and extensively studied topic in wireless sensor networks (WSNs). 
Target localization can be performed using a plethora of algorithms~\cite{Gusta05SPM,GinnakisLoc,localizationSPM} (and references therein), which exploit inter-sensor measurements like time-of-arrival (TOA), time-difference-of-arrival (TDOA), angle-of-arrival (AOA), or received signal strength (RSS). The performance of any location estimator depends not only on the algorithm but also on the placement of the anchors (sensors with known locations). Sensor placement is a key challenge in localization system design, as certain sensor constellations not only deteriorate the performance but also result in ambiguity or identifiability issues~\cite{Eusipco13chepuri}. 

The sensor placement problem can be interpreted as the problem where we divide a specific sensor area $\mathcal{S}$ into $M$ grid points and select the best subset from these grid points. Here, the selected sensors are deemed the best, if they guarantee a certain minimal accuracy on the location estimates within a specific target area $\mathcal{U}$. We consider a two-dimensional network with one target located in the target area $\mathcal{U}$ and $M$ possible sensors located at the $M$ grid points.

The absolute positions of the sensor grid points are known, hence, the considered sensors are commonly referred to as anchor nodes. Let the coordinates of the target and the $m$th anchor be denoted by the $2 \times 1$ vectors ${\boldsymbol \theta}= [\theta_1, \theta_2]^T$ and ${\bf a}_m= [a_{m,1}, a_{m,2}]^T$, respectively, where ${\boldsymbol \theta}$ is assumed to be unknown but known to be within $\mathcal{U}$.  We next illustrate the proposed framework with a number of examples all related to localization.

\subsection{Distance measurements}
Let the pairwise distance between the target and the $m$th anchor be denoted by $d_m = {\|{\boldsymbol \theta}- {\bf a}_m\|}_2$. In practice, the pairwise distances are obtained by ranging and they are generally noisy.  The range measurements generally follow an additive Gaussian non-linear model, as given by
\begin{equation}
\label{eq:dstmsr}
y_m =  d_m + n_m, \, m=1,2,\ldots,M,
\end{equation} 
where $n_m \thicksim \mathcal{N}(0, \sigma_{m}^2)$ is the noise with $\sigma_{m}^2 = \frac{\sigma^2} {d_m^{-\eta}}$. Here, $\sigma^2$ is the nominal noise variance, and $\eta$ is the path-loss exponent. 
Using \eqref{eq:fimsel}, we can now write the FIM for the localization problem as
$
{\bf C}^{-1} = {\bf F}({\bf w},{\boldsymbol \theta}) = \sum_{m=1}^{M} {w}_{,m}{\bf F}_{m}({\boldsymbol \theta}),
$ where using \eqref{eq:mFIM} we can compute
$${\bf F}_{m}({\boldsymbol \theta}) :=
\frac{({\boldsymbol \theta}-{\bf a}_m)({\boldsymbol \theta}-{\bf a}_m)^T}{\sigma_{m}^2\|{\boldsymbol \theta}-{\bf a}_m\|_2^2}.$$
\subsection{Received signal strength (RSS)}
RSS is the voltage measured by a sensor's received signal strength indicator (RSSI) circuit. RSS is often reported as the measured power. The ensemble mean received power at the $m$th sensor can be expressed as
$$
\bar{y}_m = {y_0 - 10 \eta \ln \,\frac{d_m}{d_0}}, \, m=1,2,\ldots,M,
$$
where $y_0$ is the received power (dBm) at a reference distance $d_0$. However, due to shadowing, the difference between a measured received power and its ensemble average is random. The randomness due to shadowing is typically modeled as a {\it log-normal} process, which is Gaussian if expressed in decibels~\cite{localizationSPM}. More specifically, the received power (dBm) $y_m$ at the $m$th sensor follows a Gaussian distribution, i.e., $ p(y_m;{\boldsymbol \theta}) \thicksim \mathcal{N}(\bar{y}_m, \sigma^2_{{\rm r,dB}})$. The FIM related to the $m$th measurement is then given by
$$
{\bf F}_m ({\boldsymbol \theta}) := \frac{50 \eta^2}{\sigma^2_{{\rm r,dB}}d_m^4 \ln 10} ({\boldsymbol \theta}-{\bf a}_m)({\boldsymbol \theta}-{\bf a}_m)^T.
$$
\subsection{Bearing measurements}
Another popular target localization technique is based on bearing measurements from a set of direction finding (DF) sensors~\cite{bearingonly}. The bearing measurement of the $m$th DF sensor is given by
$$
y_m = {\rm arc}\tan\left(\frac{\Delta a_{m,2}}{\Delta a_{m,1}}\right) + n_m, \, m=1,2,\ldots,M,
$$
where $\Delta a_{m,2} = \theta_2 - a_{2,m}$, $\Delta a_{m,1} = \theta_1 - a_{1,m}$, and $n_m \thicksim \mathcal{N}(0,\sigma^2_{\rm b})$ is the noise. Defining a $2 \times 2$ permutation matrix ${\bf P} = \left[\begin{array}{cc}0 & 1 \\-1 & 0\end{array}\right]$, we can then compute the FIM contribution from the $m$th DF sensor as
$$
{\bf F}_m = \frac{1}{\sigma^2_{\rm b} d_m^4} {\bf P}({\boldsymbol \theta}- {\bf a}_m)({\boldsymbol \theta}-{\bf a}_m)^T{\bf P}^T.
$$
\subsection{Energy measurements}
Another popular localization scheme relevant to field estimation, (active/passive) radar, and sonar is to estimate the location of a point source that emits or reflects energy. Suppose the sensors measure the energy generated by a point source. 
The measurements are  given as 
\begin{equation}
\label{eq:field}
y_m =   \sqrt{e} {h_m({\boldsymbol \theta})}  +  n_m, \, m=1,2,\ldots,M,
\end{equation}
where $e$ is the known energy emitted or reflected by the source, the known propagation function for some gain $\beta \ge 0$ is modeled as an isotropic exponential attenuation $h_m({\boldsymbol \theta}) = \frac{\beta}{\beta + d_m^2}$, and $n_m\thicksim \mathcal{N}(0,\sigma^2_{\rm e})$ is the noise. The FIM related to the $m$th measurement is then given by
$$
{\bf F}_m ({\boldsymbol \theta}) := \frac{4e\beta^2}{\sigma^2_{\rm e}(\beta + d_m^2)^2} ({\boldsymbol \theta}-{\bf a}_m)({\boldsymbol \theta}-{\bf a}_m)^T.
$$

\begin{myrem}[Active sensor selection]
The sensor selection problem can also be formulated for active sensing.
In active sensing, the sensors transmit probing signals (e.g., radar, sonar). The selection parameter $w_{m}$ for active sensing is a soft parameter used for {\it joint selection and resource allocation}~\cite{Eusipco13chepuri}, i.e., $w_{m} \in [0,1]$ is a resource (e.g., transmit energy) normalized to the maximum prescribed value, and hence, it is dimensionless.
The relaxed active sensor selection problem takes the same form as in \eqref{eq:l1}. In fact, minimizing the $\ell_1$-norm in active sensor selection minimizes the overall network resources (e.g., overall transmit energy).  
\end{myrem}

\begin{figure*}[!th]
\psfrag{sensors}{\small Sensor index} \psfrag{w}{\hskip-10mm \small Entries of $\hat{\bf w}$} \psfrag{k}{$\small {k}$}
\psfrag{fbestmf}{\small{$f_{\rm best}^{k} - \hat{f}$}} \psfrag{objective}{\hskip-5mm\small{$\ell_1$-norm cost}} \psfrag{pe}{\small$P_e$}
\psfrag{x}{\hskip-1cm \footnotesize x-axis coordinates [m]} \psfrag{y}{\hskip-1.5cm\footnotesize y-axis coordinates [m]}
        \centering
        \begin{subfigure}[b]{0.48\textwidth}
        \centering
                \includegraphics[width=\columnwidth,height=2.5in]{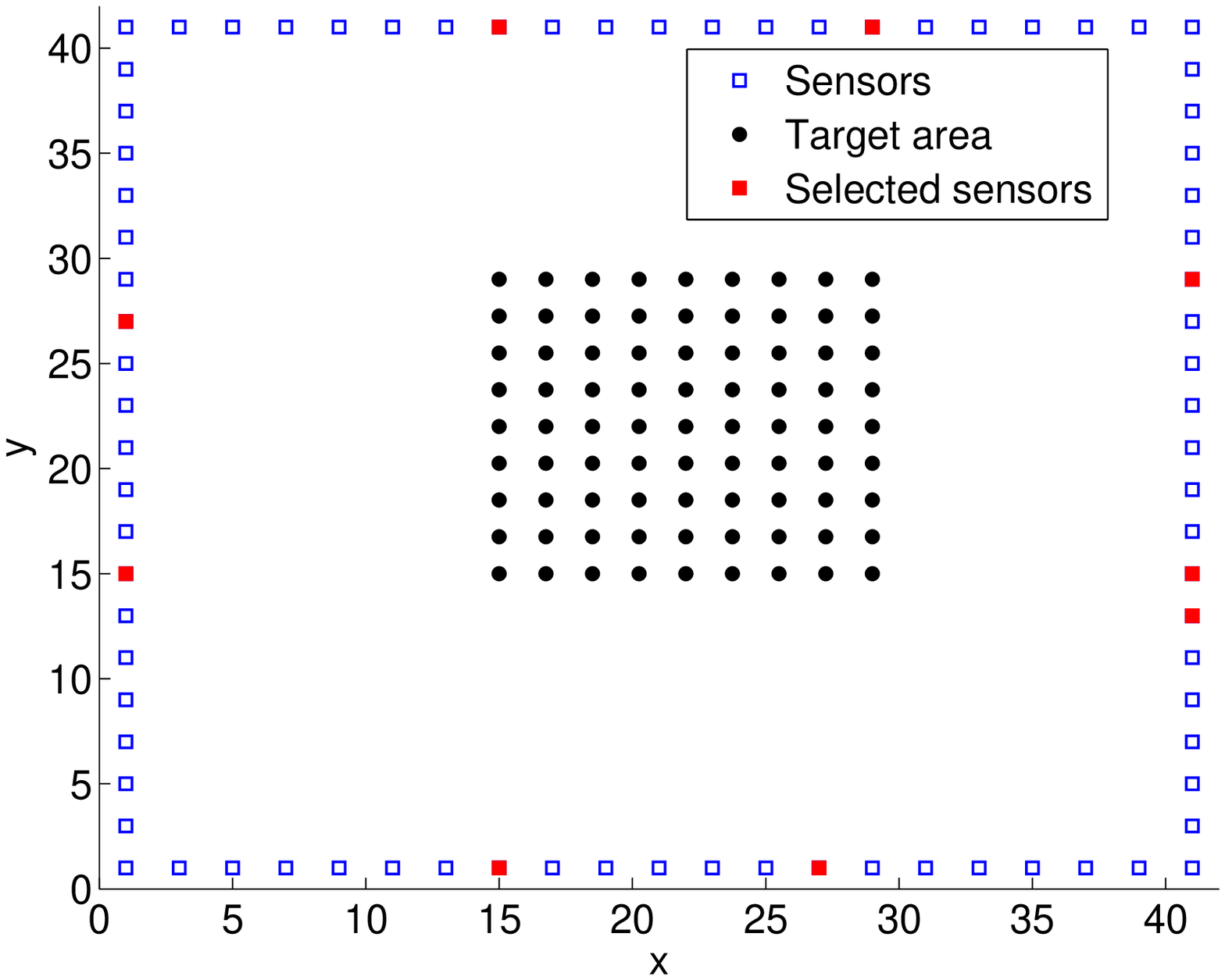}
                \caption{}
                \label{fig:sensorgridcase1}
                \end{subfigure}%
                ~
	 \begin{subfigure}[b]{0.48\textwidth}
                \centering
                \includegraphics[width=\columnwidth,height=2.5in]{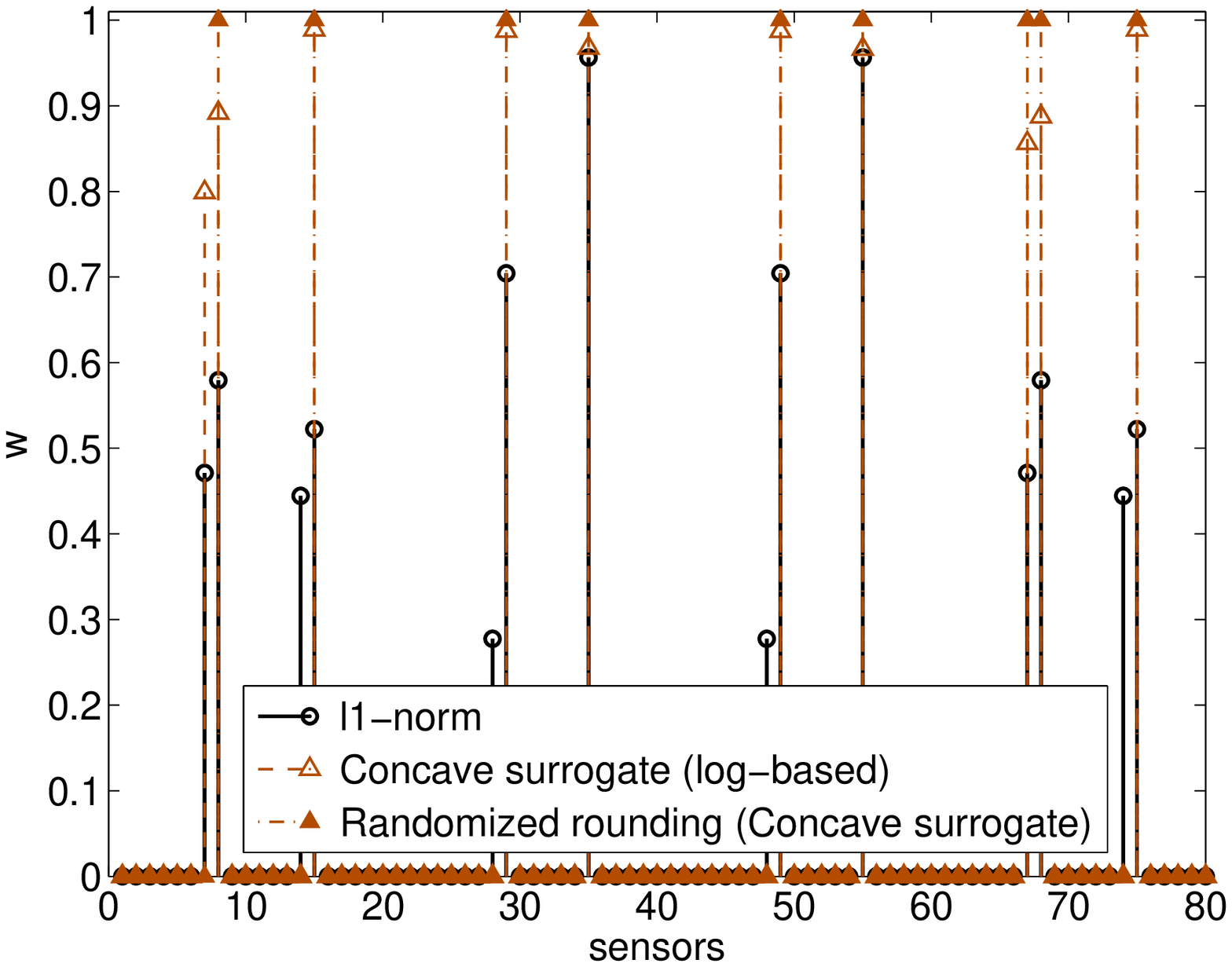}
                \caption{}
                \label{fig:sparsityenchancecase1}
        \end{subfigure}%
        \\ 
           \begin{subfigure}[b]{0.48\textwidth}
                \centering
                \includegraphics[width=\columnwidth,height=2.5in]{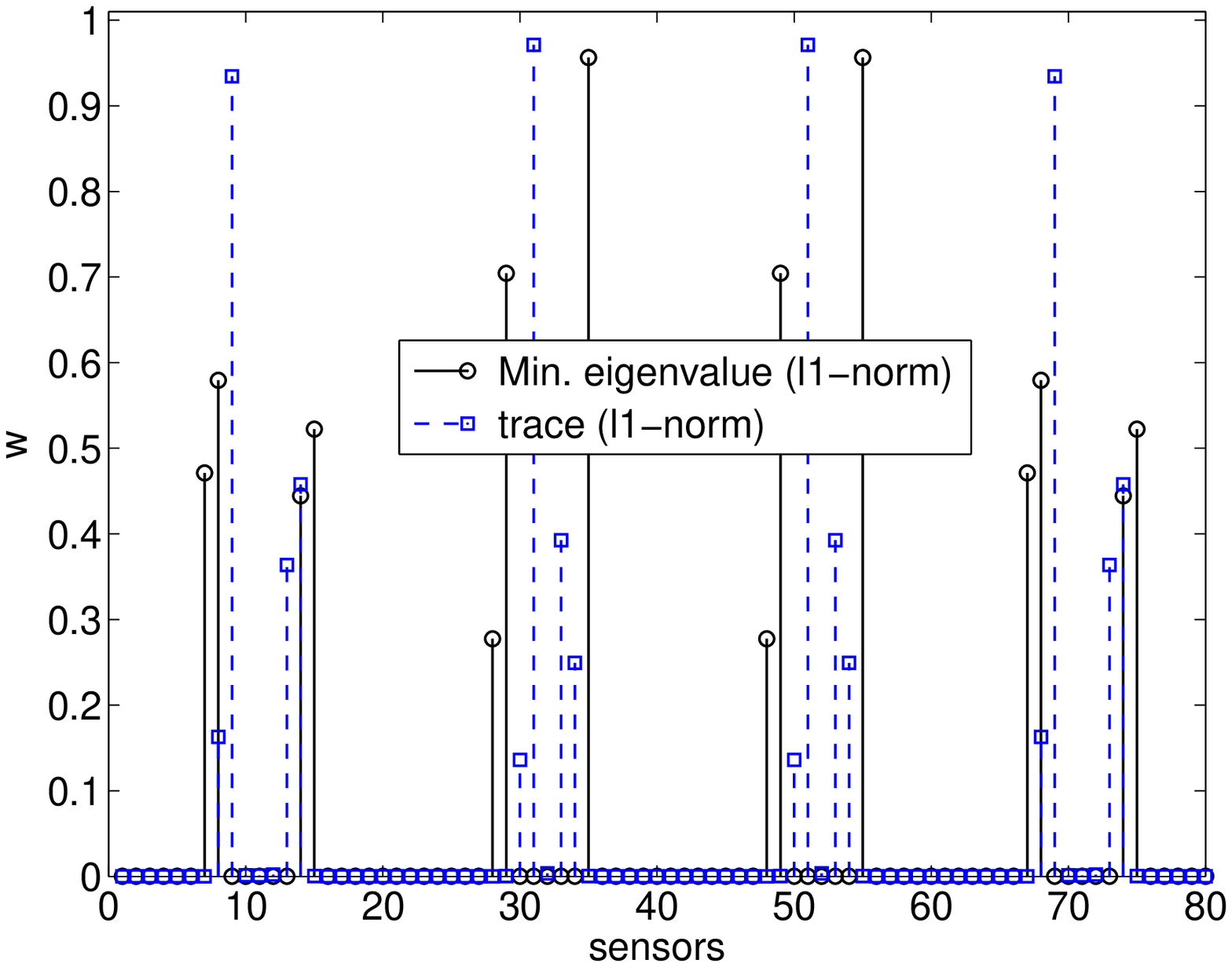}
                \caption{}
                \label{fig:tracecase1}
        \end{subfigure}
~
        \begin{subfigure}[b]{0.48\textwidth}
                \centering
                \includegraphics[width=\columnwidth,height=2.5in]{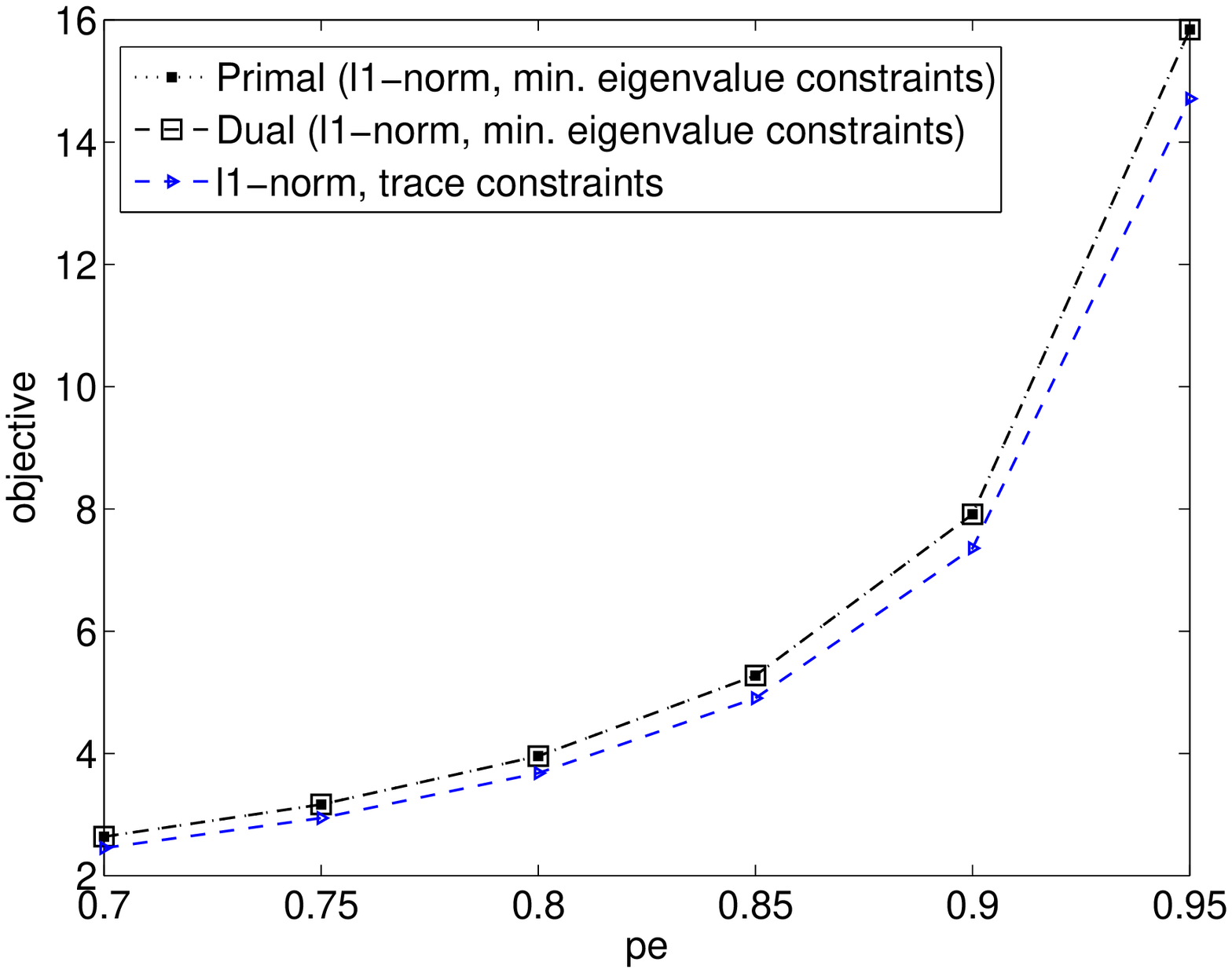}
                \caption{}
                \label{fig:objectivecase}
        \end{subfigure}
        \caption{\footnotesize{Sensor placement for target localization with $M=80$ available sensors. The thresholds are computed using $R_e = 20$~cm, and $P_e = 0.9$ (expect for (d)). (a) Selection based on sparsity-enhancing iterations with minimum eigenvalue constrains. The Boolean solution is recovered using randomized rounding. (b) Minimum eigenvalue constraints with $\ell_1$-norm and concave surrogate based relaxations. Randomized rounding is applied to the concave surrogate based solution. (c) $\ell_1$-norm based selection with the trace constraints. (d) $\ell_1$-norm cost function for different $P_e$ with $R_e=20$~cm.}}\label{fig:distancemeasurements}
\end{figure*}

\subsection{Simulations}
We apply the proposed sensor selection problem to sensor placement design for target localization. To test the proposed  algorithms, we use CVX~\cite{cvx}. CVX internally calls SeDuMi~\cite{Sturm98usingsedumi}, a MATLAB implementation of the second-order interior-point methods.

We consider the scenario shown in Fig.~\ref{fig:sensorgridcase1} with $M=80$ sensors to illustrate the sensor selection problem. Recall that the problem here is to choose the best sensor positions out of $M=80$ available ones, such that a certain specified localization accuracy is achieved. The domain $\mathcal{U}$ for this example will be the target (or surveillance) area where the target resides, and to avoid having infinitely many constraints the area $\mathcal{U}$ consists of grid points at a certain resolution. We grid the target area of $15 \times 15~{\rm m}^2$ uniformly with a resolution of $1.75$~m along both horizontal and vertical directions as shown in Fig.~\ref{fig:sensorgridcase1}. 

The original non-convex sensor selection problem is relaxed to an $\ell_1$-norm optimization problem. Alternatively, a concave surrogate function can be used to enhance the sparsity. The optimization problem with the concave surrogate cost function is iteratively solved by affinely scaling the objective based on the solution from the previous iteration. For the sparsity-enhancing iterative Algorithm~\ref{alg:iterativealgo}, we use $i_{\rm max}=10$ and $\delta = 10^{-8}$. The number of candidates used in the randomized rounding Algorithm~\ref{alg:randomizedround} is $L=100$. As observed in the simulations, a solution is typically found in the first batch itself, and a few tens of candidate entries are sufficient. 
We use the following parameters for the simulations:  $\eta=2$, $\sigma^2  = 2 \times 10^{-5}$, $\sigma_{\rm b}^2 = 2 \times 10^{-5}$~square-degrees, $\sigma_{{\rm r, dB}} = 2$~dB, $\sigma_{\rm e}^2 = -20$~dBm, $e=1$, $\beta=1$, and $P_e=0.9$ (except in Fig.~\ref{fig:objectivecase}).

Fig.~\ref{fig:distancemeasurements} shows the sensor selection for the distance (range) measurement model. The thresholds are computed with $R_e = 20$~cm and $P_e =0.9$. The selection shown in Fig.~\ref{fig:sensorgridcase1} is based on Algorithm~\ref{alg:iterativealgo} with randomized rounding to recover the approximate Boolean solution. The selection results based on the $\ell_1$-norm cost with the minimum eigenvalue constraint is shown in Fig.~\ref{fig:sparsityenchancecase1}. Fig.~\ref{fig:sparsityenchancecase1} also shows that the solution based on the concave surrogate cost function with the minimum eigenvalue constraint leads to a sparser solution.  The selection results based on the trace constraint obtained by solving \eqref{eq:l1trace} are illustrated in Fig.~\ref{fig:tracecase1}. The sensors from the same region (close to the red filled boxes in Fig.~\ref{fig:sensorgridcase1}) are selected with both constraints. Fig.~\ref{fig:objectivecase} shows a {\it zero-duality} gap (gap between the cost of the primal problem in \eqref{eq:l1} and the dual problem in \eqref{eq:dual}) for different values of $P_e$. Larger values of $P_e$ result in a larger ${\|{\bf w}\|}_1$, and subsequently more sensors are selected. The sufficient trace constraint has a larger feasible set compared to the stronger sufficient minimum eigenvalue constraint. As a result, for the considered scenario, the minimum eigenvalue constraint leads to a slightly larger $\ell_1$-norm compared to the trace constraint.  

\begin{figure*}[!t]
\psfrag{sensors}{\small Sensor index} \psfrag{w}{\hskip-10mm \small Entries of $\hat{\bf w}$} \psfrag{k}{$\small {k}$}
\psfrag{fbestmf}{\small{$(f_{\rm best}^{k} - {f}_{\rm opt})/{f}_{\rm opt}$}} \psfrag{epsilon10000000000000000}{\small{$\epsilon=f_{\rm best}^{k} + 10/ (10+k)$}} \psfrag{pe}{\small$P_e$}
\psfrag{x}{\hskip-1cm \footnotesize x-axis coordinates [m]} \psfrag{y}{\hskip-1.5cm\footnotesize y-axis coordinates [m]}
        \centering
        	\begin{subfigure}[b]{0.48\textwidth}
                \centering
                \includegraphics[width=\columnwidth,height=2.5in]{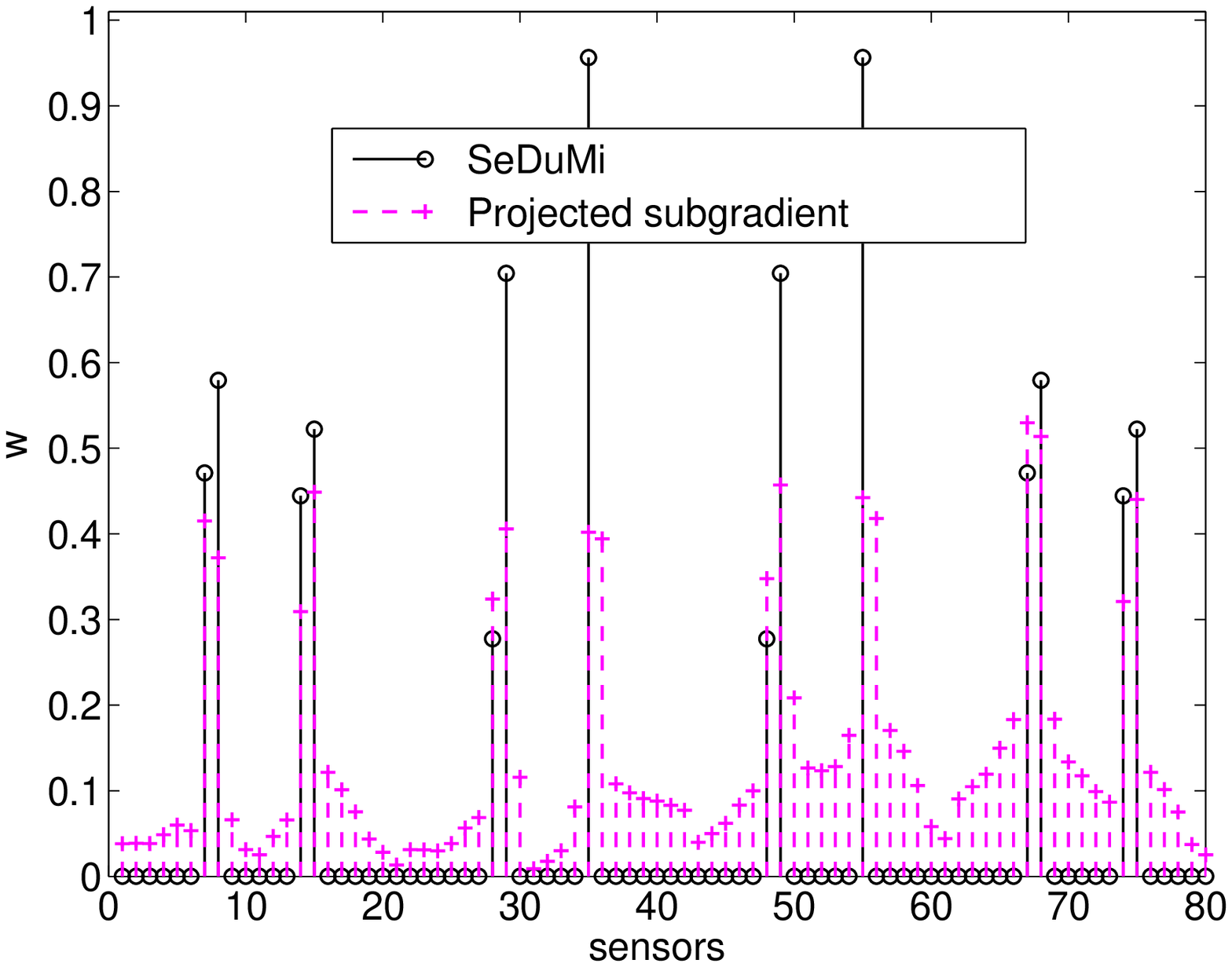}
                \caption{}
                \label{fig:projsubgradient}
        \end{subfigure} 
        ~
       \begin{subfigure}[b]{0.48\textwidth}
 	\centering
                \includegraphics[width=\columnwidth,height=2.5in]{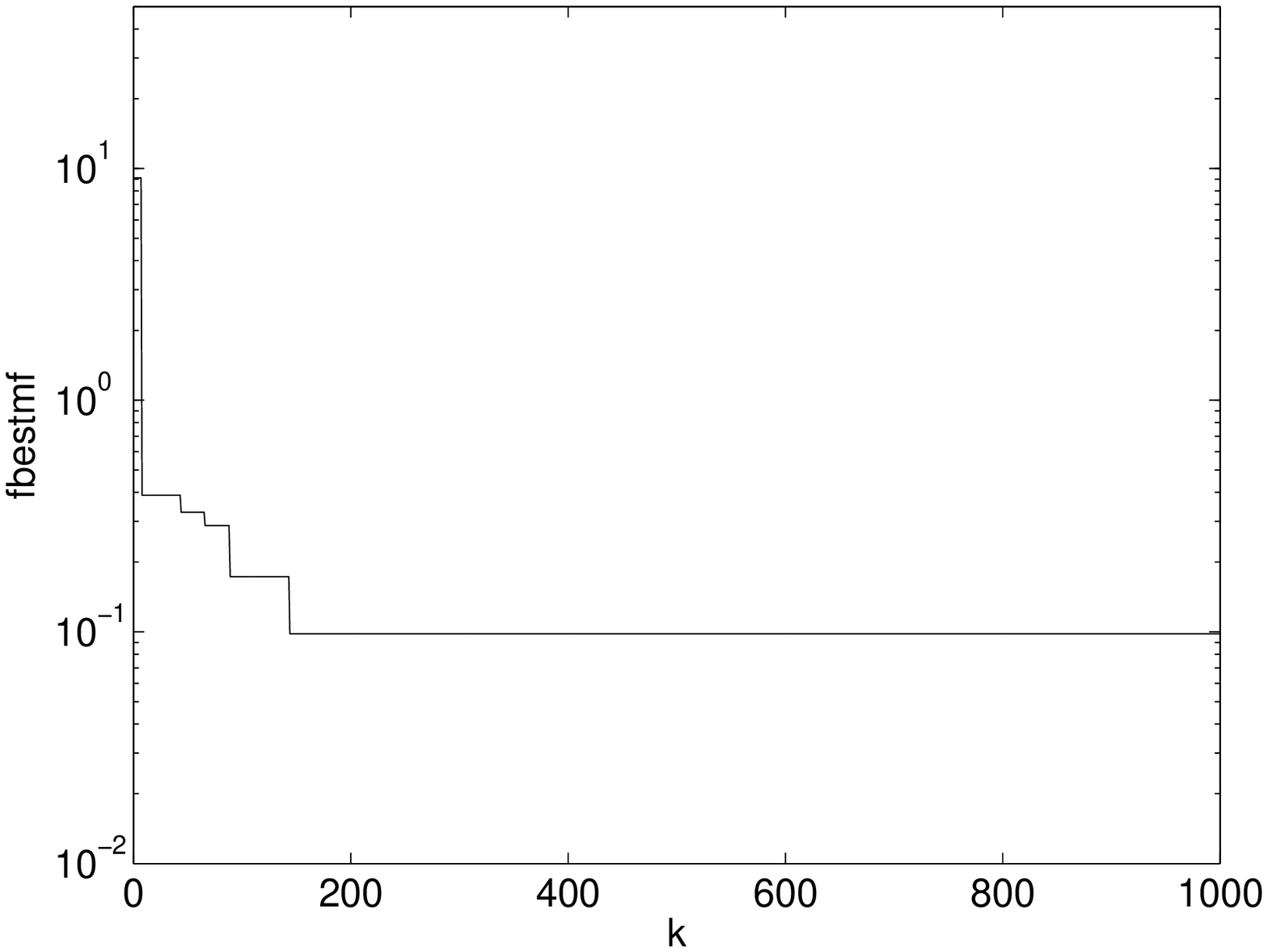}
                \caption{}
                \label{fig:projsubconvergence}
        \end{subfigure}%
        \caption{\footnotesize{(a) The projected subgradient algorithm used to solve \eqref{eq:l1}. (b) Performance of the projected subgradient algorithm. The thresholds are computed using $R_e = 20$~cm, and $P_e = 0.9$.}}\label{fig:case1}
\end{figure*}

The optimization problem \eqref{eq:l1} is also solved using the projected subgradient method summarized in Algorithm~\ref{alg:projsubalgo} with $k_{\rm max} = 1000$ iterations. The solution of the projected subgradient is shown in Fig.~\ref{fig:projsubgradient}. 
The performance of the projected subgradient algorithm is compared to the solution of the interior-point methods (implemented using SeDuMi) denoted by ${f}_{\rm opt}$ (obtained via SeDuMi), i.e., 
$(f_{\rm best}^{k} - {f}_{\rm opt})/{f}_{\rm opt}$ is shown in Fig.~\ref{fig:projsubconvergence}. Even though the convergence of the projected subgradient algorithm is very slow, the estimated support after a few hundred iterations can be used along with randomized rounding to further refine the solution. The computation time on the same computer for the projected subgradient algorithm that solves \eqref{eq:l1} is around $8.84$ seconds for $1000$ iterations while SeDuMi takes around $4.03$ seconds to solve the SDP problem in \eqref{eq:l1}. 

\begin{figure*}
\psfrag{x}{\hskip-1cm \footnotesize x-axis coordinates [m]} \psfrag{y}{\hskip-1cm\footnotesize y-axis coordinates [m]}
\psfrag{sensors} {\footnotesize Sensors} \psfrag{w}{\hskip-1.3cm \footnotesize Entries of $\hat{\bf w}_{\rm a}$ (normalized to $e$)}
\psfrag{objective}{\hskip-5mm\small{$\ell_1$-norm cost}} \psfrag{Re [cm]}{\hskip-2mm\small$R_e$~[cm]}
\psfrag{sqrt(1-Pe)Rep2}{\tiny$\sqrt{(1-P_e)R_e^2}$} \psfrag{RMSE [cm]}{\hskip-2mm\small RMSE~[cm]}
\psfrag{sensor index}{\hskip-2mm\small Sensor index}
        \centering
               \begin{subfigure}[b]{0.48\textwidth}
                \centering
                \includegraphics[width=\columnwidth,height=2.5in]{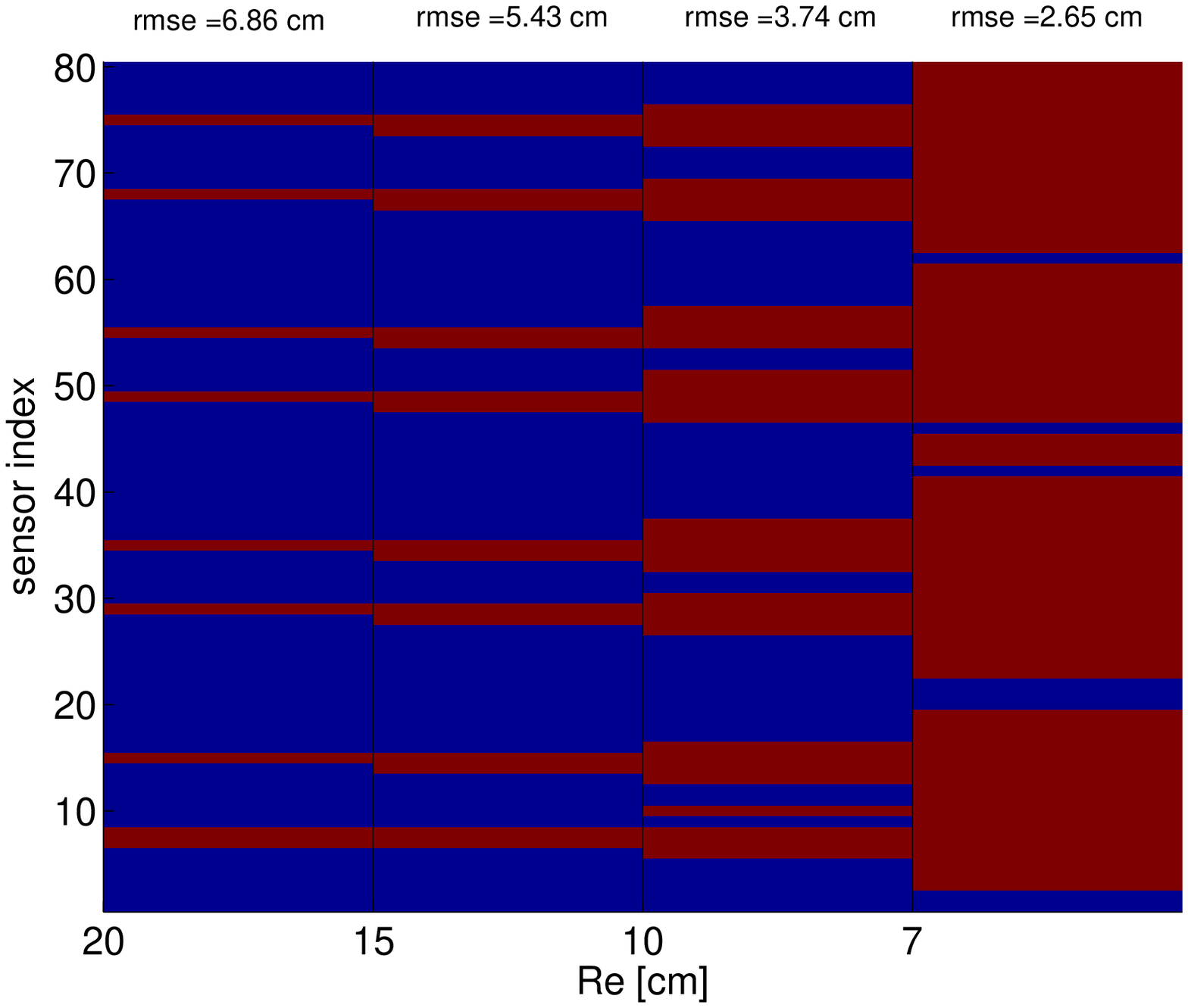}
                \caption{}
                \label{fig:solutionpath}
        \end{subfigure}
        ~%
                \begin{subfigure}[b]{0.48\textwidth}
                \centering
                \includegraphics[width=\columnwidth,height=2.5in]{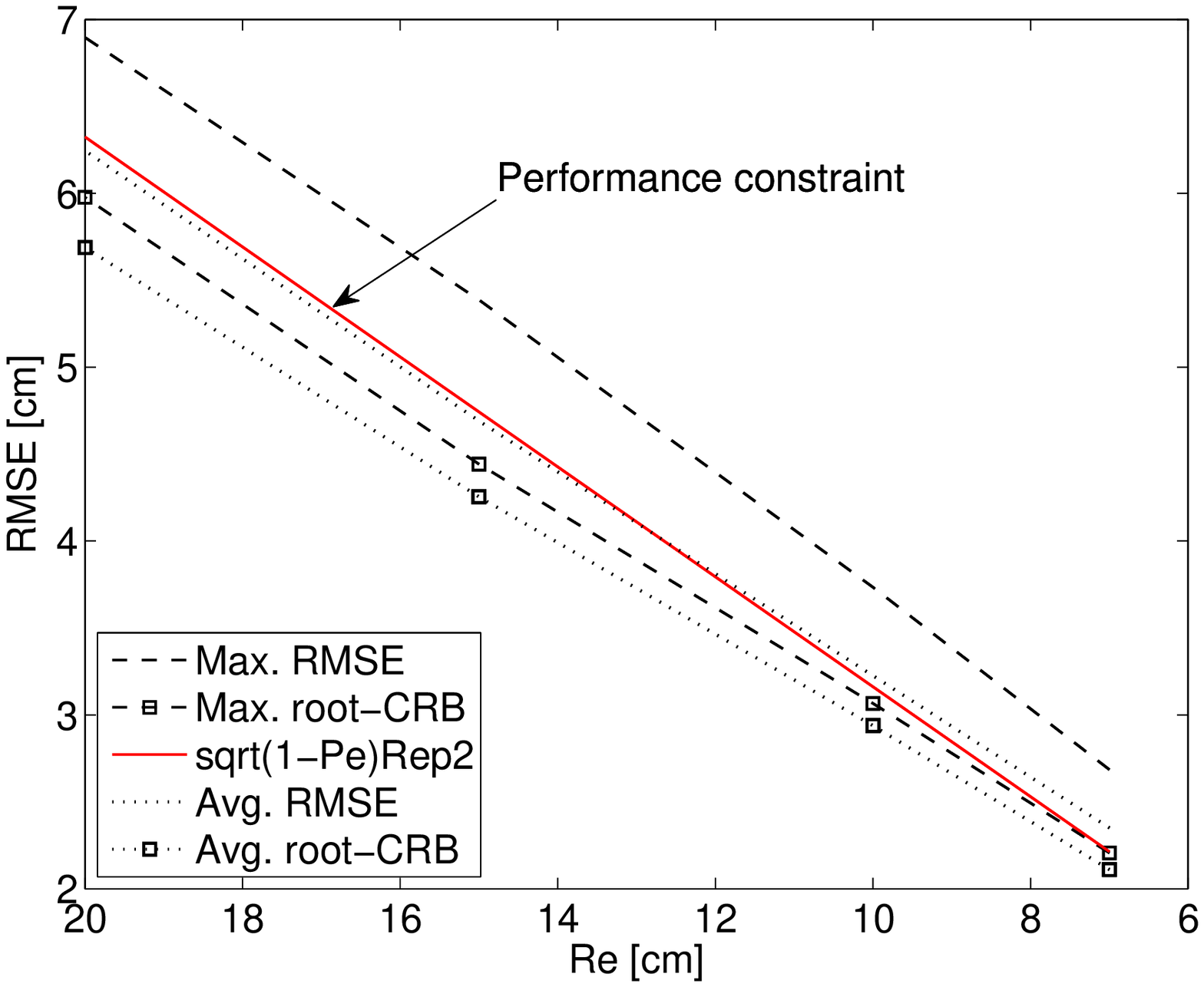}
                \caption{}
                \label{fig:rmsecrb}
        \end{subfigure}
        \caption{\footnotesize{(a) Solution path of the sensors selected for different values of $R_e$ and $P_e =0.9$. Maximum RMSE based on selected sensors can be seen on the top of this plot. (b) Maximum and average RMSE of the location estimates based on Gauss-Newton's method, the corresponding maximum and average root-CRB, and the performance constraint in \eqref{eq:chebapp} for different values of $R_e$, and $P_e = 0.9$.}}\label{fig:case2}
\end{figure*}
A practical estimator does not meet the CRB in some cases (for instance at low SNRs or finite data records). Therefore, the sensors obtained with a specific $R_e$ would lead to an underestimate of the desired MSE. We can account for this gap by choosing $R_e$ appropriately. To this end, we give the entire solution path of the selected sensors for different values of $R_e$ in Fig.~\ref{fig:solutionpath}. The solution path can be efficiently computed by increasing $R_e$. The sensors corresponding to some $R_e$ can then be used to meet the desired MSE requirement. The non-linear model in \eqref{eq:dstmsr} is solved in the least-squares sense iteratively using Gauss-Newton's method with $10$ iterations~\cite{SKayestimation}. The maximum root-MSE (RMSE), maximum root-CRB, average RMSE, and average CRB of the location estimates of a target within the target area using the selected sensors (as shown in the solution path) for different values of $R_e$ are shown in Fig.~\ref{fig:rmsecrb}. For the considered scenario, both the maximum and average root-CRB satisfy the performance constraint which is given by the inequality in \eqref{eq:chebapp}. The performance constraint is shown as a red solid line in Fig.~\ref{fig:rmsecrb}. The maximum RMSE does not satisfy the accuracy requirement specified by a certain $R_e$, and this can be corrected by using an appropriate (lower) $R_e$. Moreover, for the considered scenario, the gap between the average RMSE and the performance constraint is still reasonable. We also show the maximum RMSE on top of Fig.~\ref{fig:solutionpath}. 

\begin{figure*}
\psfrag{sensors}{\small Sensor index} \psfrag{w}{\hskip-10mm \small Entries of $\hat{\bf w}$} \psfrag{k}{$\small {k}$}
\psfrag{fbestmf}{\small{$(f_{\rm best}^{k} - {f}_{\rm opt})/{f}_{\rm opt}$}} \psfrag{epsilon1000}{\small{$f_{\rm best}^{k} + 10/ (10+k)$}} \psfrag{pe}{\small$P_e$}
\psfrag{x}{\hskip-1cm \footnotesize x-axis coordinates [m]} \psfrag{y}{\hskip-1.5cm\footnotesize y-axis coordinates [m]}
        \centering
        	\begin{subfigure}[b]{0.48\textwidth}
                \centering
                \includegraphics[width=\columnwidth,height=2.5in]{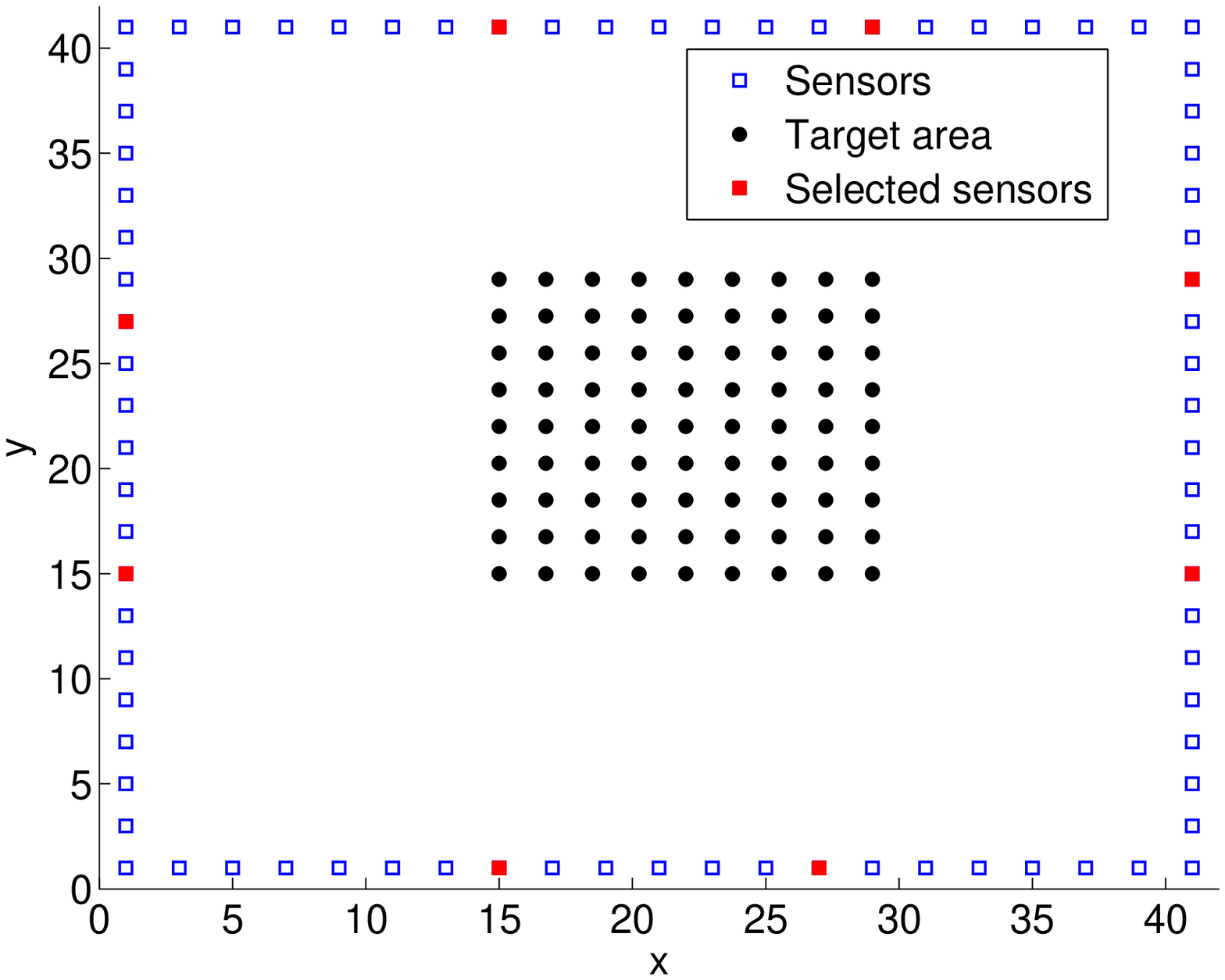}
                \caption{}
                \label{fig:BearingSentopology}
        \end{subfigure} 
        ~
       \begin{subfigure}[b]{0.48\textwidth}
 	\centering
                \includegraphics[width=\columnwidth,height=2.5in]{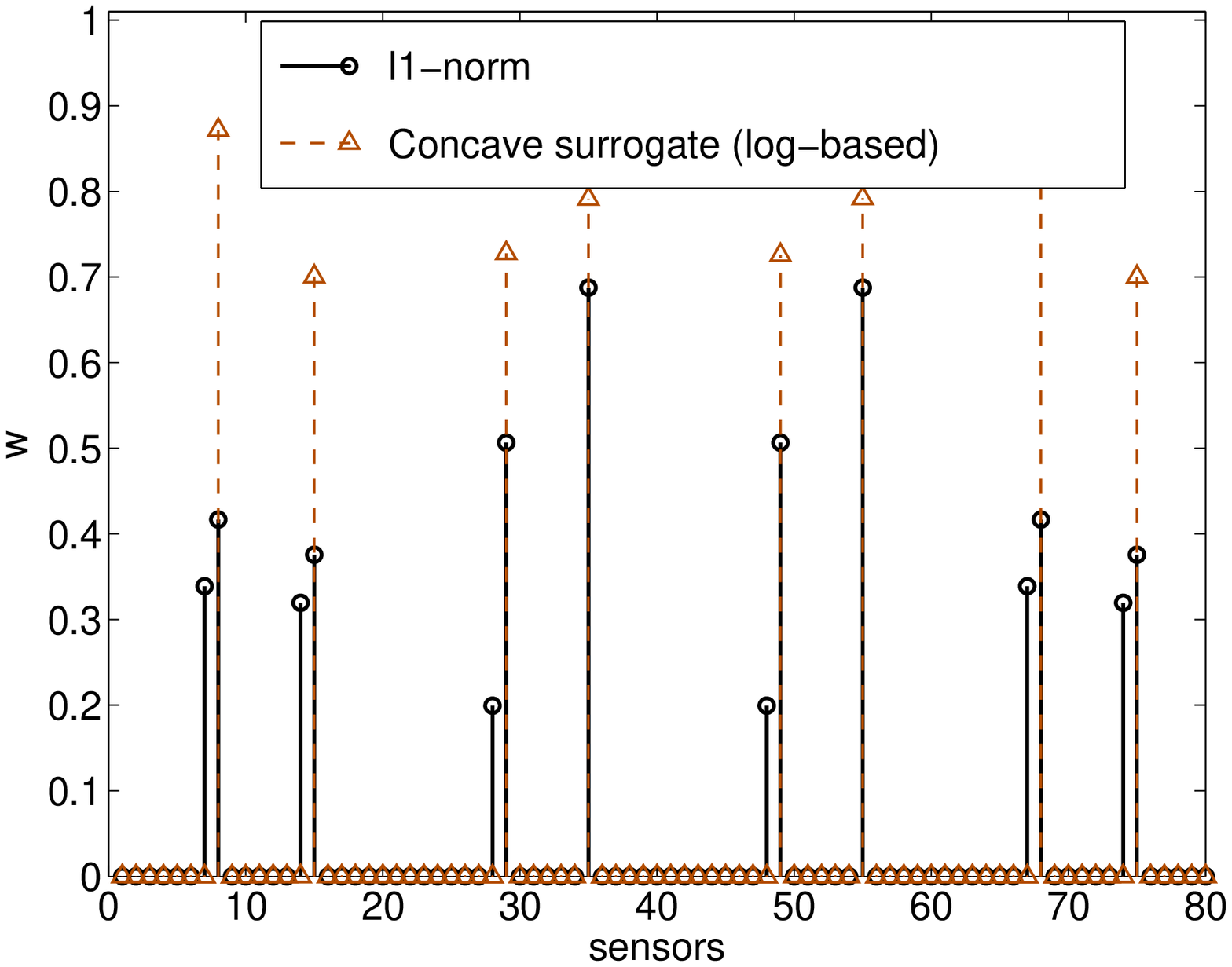}
                \caption{}
                \label{fig:BearingSensel}
        \end{subfigure}%
        \caption{\footnotesize{(a) Sensor selection based on bearing measurements with $M=80$ available sensors. (b) Sensor selection solved with minimum eigenvalue constraints using $\ell_1$-norm and log-based heuristics. The thresholds are computed using $R_e = 25$~cm, and $P_e = 0.9$. The noise variance is $\sigma_{\rm b}^2 = 2 \times 10^{-5}$~square-degrees.}}\label{fig:Bearing}
\end{figure*}

\begin{figure*}
\psfrag{sensors}{\small Sensor index} \psfrag{w}{\hskip-10mm \small Entries of $\hat{\bf w}$} \psfrag{k}{$\small {k}$}
\psfrag{fbestmf}{\small{$(f_{\rm best}^{k} - {f}_{\rm opt})/{f}_{\rm opt}$}} \psfrag{epsilon1000}{\small{$f_{\rm best}^{k} + 10/ (10+k)$}} \psfrag{pe}{\small$P_e$}
\psfrag{x}{\hskip-1cm \footnotesize x-axis coordinates [m]} \psfrag{y}{\hskip-1.5cm\footnotesize y-axis coordinates [m]}
        \centering
        	\begin{subfigure}[b]{0.48\textwidth}
                \centering
                \includegraphics[width=\columnwidth,height=2.5in]{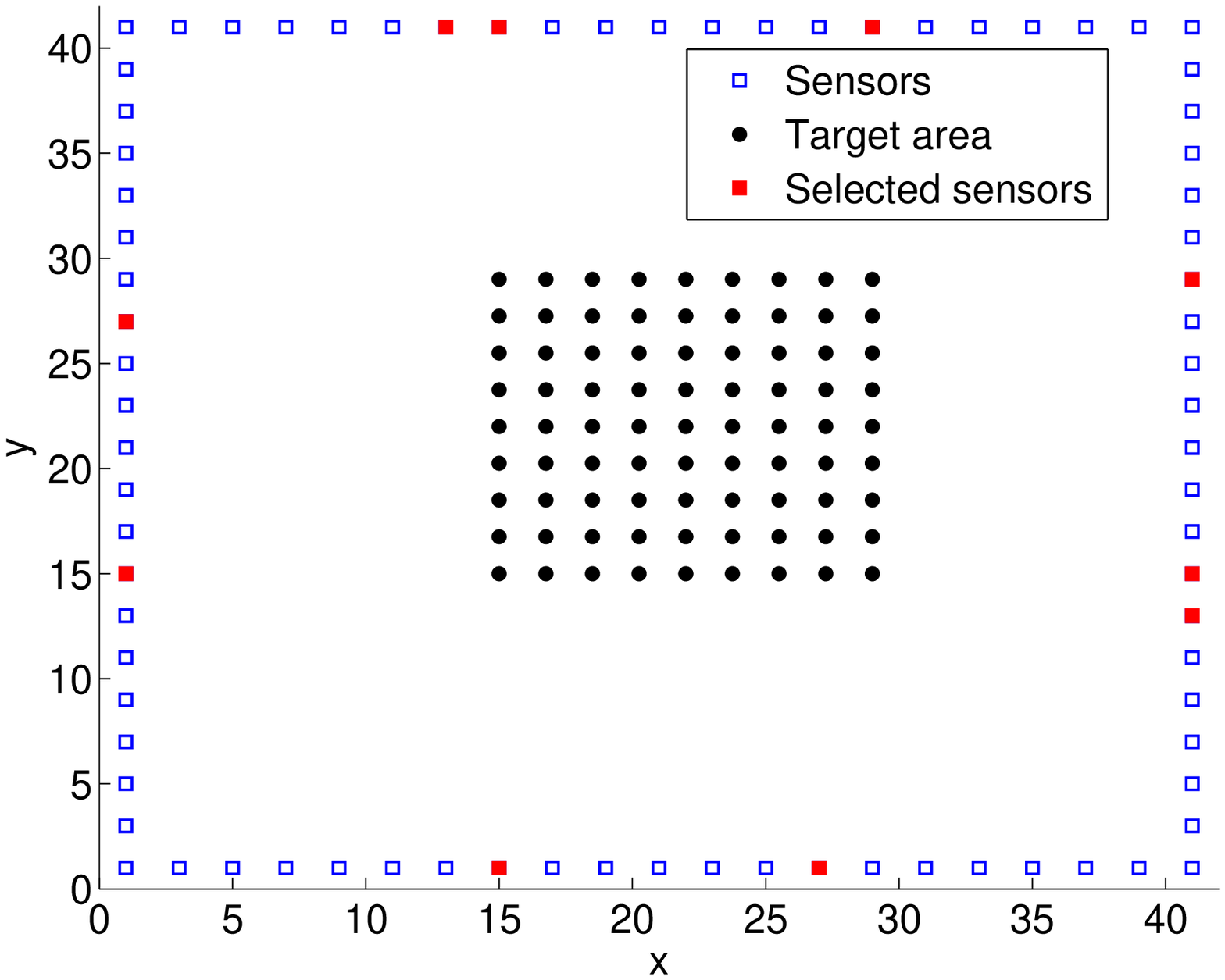}
                \caption{}
                \label{fig:PathlossSentopology}
        \end{subfigure} 
        ~
       \begin{subfigure}[b]{0.48\textwidth}
 	\centering
                \includegraphics[width=\columnwidth,height=2.5in]{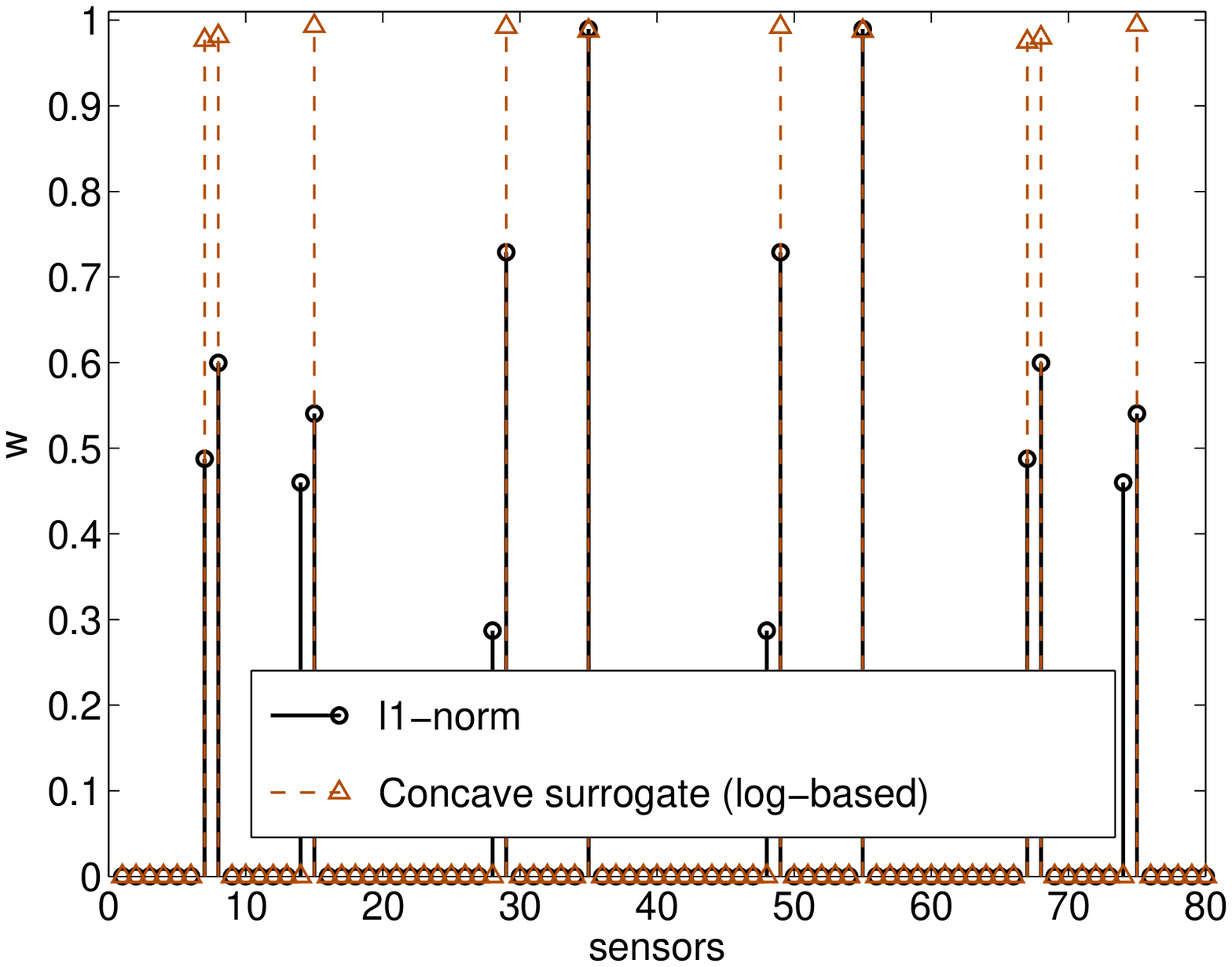}
                \caption{}
                \label{fig:PathlossSensel}
        \end{subfigure}%
        \caption{\footnotesize{(a) Sensor selection based on RSS with $M=80$ available sensors. (b) Sensor selection solved with minimum eigenvalue constraints using $\ell_1$-norm and log-based heuristics. The thresholds are computed using $R_e = 5$~m, and $P_e = 0.9$. We use $\sigma_{{\rm r,dB}} = 2$~dB.}}\label{fig:pathloss}
\end{figure*}

\begin{figure*}
\psfrag{sensors}{\small Sensor index} \psfrag{w}{\hskip-10mm \small Entries of $\hat{\bf w}$} \psfrag{k}{$\small {k}$}
\psfrag{fbestmf}{\small{$(f_{\rm best}^{k} - {f}_{\rm opt})/{f}_{\rm opt}$}} \psfrag{epsilon1000}{\small{$f_{\rm best}^{k} + 10/ (10+k)$}} \psfrag{pe}{\small$P_e$}
\psfrag{x}{\hskip-1cm \footnotesize x-axis coordinates [m]} \psfrag{y}{\hskip-1.5cm\footnotesize y-axis coordinates [m]}
        \centering
        	\begin{subfigure}[b]{0.48\textwidth}
                \centering
                \includegraphics[width=\columnwidth,height=2.5in]{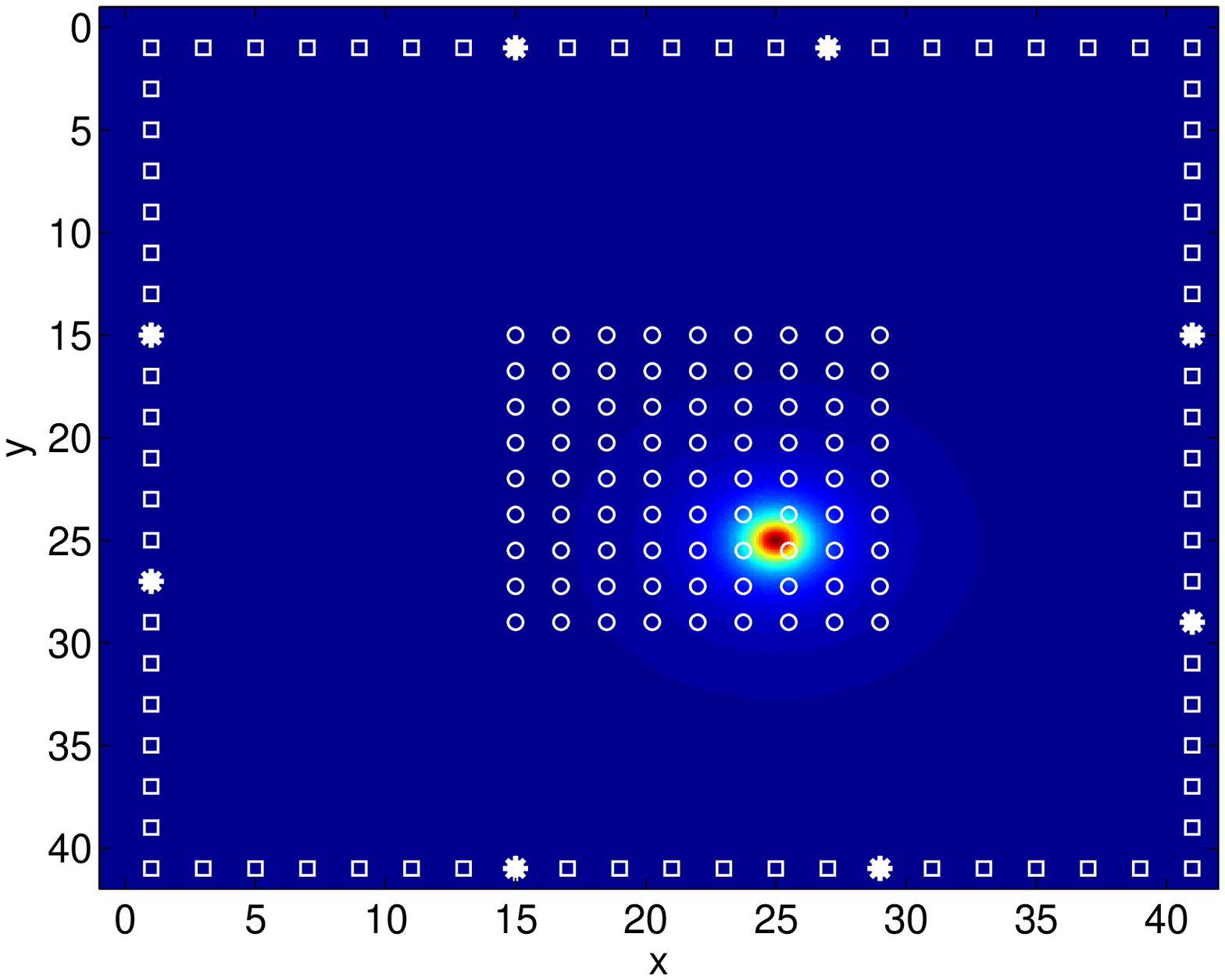}
                \caption{}
                \label{fig:RssSentopology}
        \end{subfigure} 
        ~
       \begin{subfigure}[b]{0.48\textwidth}
 	\centering
                \includegraphics[width=\columnwidth,height=2.5in]{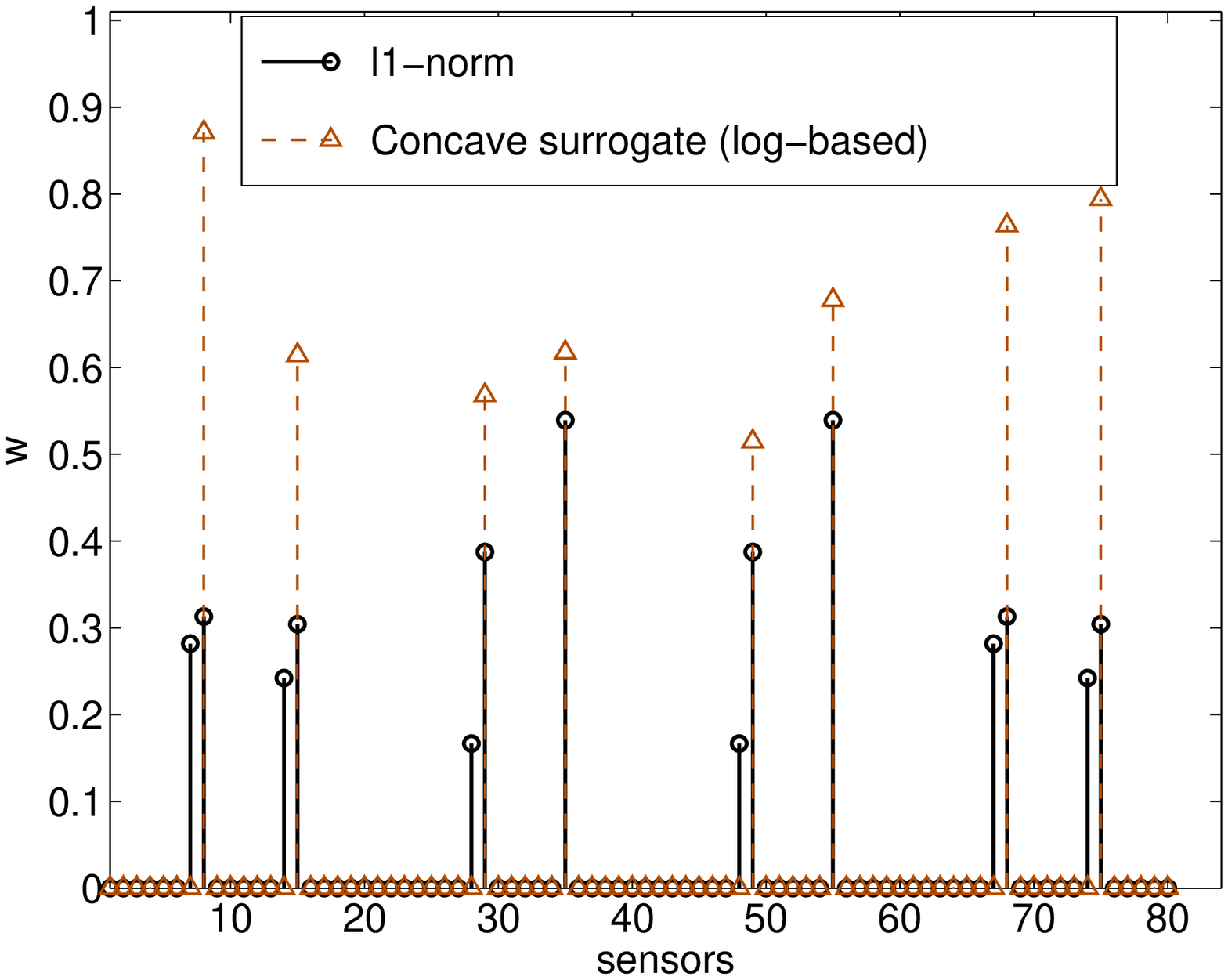}
                \caption{}
        \end{subfigure}%
        \caption{\footnotesize{Sensor selection based on energy measurements. (a) Illustration of a field generated by a unit amplitude point source at location ${\boldsymbol \theta} = [25,25]^T$~m according to \eqref{eq:field}. Out of $M=80$ available sensors $(\square)$, 16 sensors indicated by $(\ast)$ are selected. The source domain is indicated by $(\circ)$. (b) Sensor selection solved with minimum eigenvalue constraints using $\ell_1$-norm and log-based heuristics. The thresholds are computed using $R_e = 10$~cm, and $P_e = 0.9$. We use $e=1$, $\beta = 1$ and $\sigma_{\rm e}^2 = 10^{-5}$.}}\label{fig:field estimation}\label{fig:FeildSensel}
\end{figure*}

The proposed framework is very general, and can be applied to a variety of data models as long as (a1) and (a2) are valid. To illustrate this we next consider a few more measurement models. The sensor selection based on bearing measurements is illustrated in Fig.~\ref{fig:Bearing}. Here, we use $R_e = 25$~cm, and $P_e = 0.9$. The selection results for the RSS based measurement model is shown in Fig.~\ref{fig:pathloss}. We use $\sigma_{{\rm r,dB}}^2 = 2$~dB, $P_e = 0.9$, and $R_e = 5$~m. Sensor selection results based on energy measurements are shown in Fig.~\ref{fig:FeildSensel}, where we use $R_e=10$~cm, and $P_e=0.9$. An illustration of the field generated by a point source at location $[25,25]^T$~m with unit amplitude is also shown here. 

The FIM for all the considered measurement models has a common structure, and it decreases as the distance $d_m$ increases. However, the rate at which it decreases is different for different models. Anyway, as a result of this decrease, the optimization problem leads to a sensor selection that is close to the target area (in the Euclidean distance sense) for all the considered models.

\section{Conclusions}\label{sec:conclusion}
Sensor selection is an important design problem in sensor networks. The sensor selection problem can be described as the problem of selecting the best subset of sensors that guarantees a certain specified performance measure. The sensor selection enables deployment of the sensors with guarantees on the resulting estimation accuracy. This also minimizes the hardware, communications, and resulting processing costs in large-scale networks. We focus on observations that follow a non-linear model. The proposed framework is valid as long as the observations are independent, and their pdfs are regular. We use a number of functions related to the FIM as a performance measure. The original nonconvex optimization problem is relaxed using convex relaxation techniques which can then be efficiently solved in polynomial time. To handle large-scale problems, we have also presented a projected subgradient algorithm. This also enables easy distributed  implementations. The proposed framework is applied to sensor placement design for a number of different models related to localization.

\appendices

\section{Performance thresholds}\label{app:RSPconstraints}
\subsection*{Trace and minimum eigenvalue constraints}
We can relate the accuracy requirement and the CRB using Chebyshev's inequality~\cite{cover2012elements}
$$
\mathrm{Pr}(\|\boldsymbol \epsilon \|_2 \geq R_e) \leq {\rm tr}\{{\bf C}\}/R_e^2
$$ 
which can be equivalently expressed as $\mathrm{Pr}(\|\boldsymbol \epsilon \|_2 \leq R_e) \leq 1- {\rm tr}\{{\bf C}\}/R_e^2$. Combining this inequality together with $\mathrm{Pr}(\|\boldsymbol \varepsilon \|_2 \leq R_e) \geq P_e$ in \eqref{eq:accuracyrequired} results in the following sufficient condition
\begin{equation}
\label{eq:chebapp}
{\rm tr}\{{\bf C}\} \leq \lambda_{\rm tr} = (1- P_e)R_e^2.
\end{equation}
Each eigenvalue of ${\bf C}^{-1}$ is greater than $\lambda_{\rm min}({\bf F})$, and as a result, ${\rm tr}\{{\bf C}\} \leq N \lambda_{\rm min}^{-1}({\bf F})$. Hence, a stronger sufficient condition (with a smaller feasible set) is $N \lambda_{\rm min}^{-1}({\bf F}) \leq (1-P_e)R_e^2$, or equivalently~\cite{Tao09TSP}
$$\lambda_{\rm min}({\bf F}) \geq  \lambda_{\rm eig}=\frac{N}{R_e^2} \left(\frac{1}{1-P_e}\right).$$

\subsection*{Determinant constraint}
The determinant constraint is related to the volume or the mean radius of the confidence ellipsoid that contains ${\boldsymbol \epsilon} = {\boldsymbol \theta} - \hat{\boldsymbol \theta}$ with probability ${P}_e$. Such a confidence ellipsoid can be expressed as
$$
\mathcal{E} = \{ {\boldsymbol \epsilon} \, | \,  {\boldsymbol \epsilon}^T {\bf F}^{-1}{\boldsymbol \epsilon} \leq \xi  \},
$$ 
where  $\xi$ is a constant that depends on $P_e$. Assuming ${\bf F}$ has ordered eigenvalues $\lambda_{\rm max} \geq \lambda_2 \cdots \geq \lambda_{\rm min}$, the length of the $n$th semi-axis of the ellipsoid $\mathcal{E}$ will be $\sqrt{{\xi}/\lambda_{n}}$. The geometric mean radius of the confidence ellipsoid $\mathcal{E}$ given by 
$$ 
\bar{R}_e = {\sqrt{\xi}}/ ({\rm det} \{{\bf F}\})^{1/2N}, 
$$
gives a quantitative measure of how informative the observations are. For the estimates to be within the confidence ellipsoid $\mathcal{E}$, we use the constraint  
$$
 \ln \,{\rm det} \{{\bf F}\} \geq  2N \ln \, \frac{\sqrt{\xi}}{\bar{R}_e} = \lambda_{\rm det},
$$
where $\bar{R}_e$ and $\sqrt{\xi}$ specify the required accuracy, and are assumed to be known. A typical choice for $\xi$ is constant chi-squared values, i.e., $\xi = F^{-1}_{\mathcal{X}^2_N}(P_e)$.
Here, $F^{-1}_{\mathcal{X}^2_N}$ is the cumulative distribution function of a chi-squared random variable with $N$ degrees of freedom. This performance measure is related to the D-optimality.
\section{Derivation of the dual problem}\label{app:dual}
Consider the optimization problem \eqref{eq:l1} as the primal problem.
To this problem, we then associate the following dual variables or Lagrangian multipliers: ${\bf Z} \in  \mathbb{S}^{DN}$ with the LMI constraint; $\nu_m \in \mathbb{R}$ and $\mu_m  \in \mathbb{R}$ with the $w_m \geq 0$ and $w_m \leq 1$ constraints, respectively. The Lagrangian is
\begin{equation*}
\begin{aligned}
{\mathcal L}({\bf w},{\bf Z}, {\boldsymbol \mu}, {\boldsymbol \nu}) &= {\bf 1}_M^T{\bf w} - \sum_{m=1}^M w_m  {\rm tr}\{{\bf F}_m{\bf Z}\} - {\rm tr}\{-\lambda_{\rm eig}{\bf Z}\}\\
&\hskip10mm + {\boldsymbol \mu}^T({\bf w} - {\bf 1}_M) - {\boldsymbol \nu}^T{\bf w} \\
&= \sum_{m=1}^{M} w_m (1- {\rm tr}\{{\bf F}_m{\bf Z}\} + \mu_m - \nu_m) \\  
&\hskip10mm - {\bf 1}_M^T{\boldsymbol \mu} - {\rm tr}\{-\lambda_{\rm eig}{\bf Z}\}.
\end{aligned}
\end{equation*}

The Lagrange dual function $$\phi({\bf Z}, {\boldsymbol \mu}, {\boldsymbol \nu})= \inf_{\bf w} {\mathcal L}({\bf w},{\bf Z}, {\boldsymbol \mu}, {\boldsymbol \nu})$$ is given as \eqref{eq:Ldual}.
The dual problem which is also an SDP can therefore be expressed as
\begin{equation*}
\begin{aligned} 
\hskip0.5mm &\argmax_{{\bf Z}, \, {\boldsymbol \mu}, {\boldsymbol \nu}} \quad \lambda_{\rm eig}\,{\rm tr}\{{\bf Z}\}-{\bf 1}_M^T {\boldsymbol \mu}
\\
&\hskip1mm{\rm s.t.} \quad {\rm tr}\{{\bf F}_m{\bf Z}\} + \nu_m = 1 + \mu_m, \, m=1,2,\ldots,M, \\ 
&\hskip10mm {\bf Z} \succeq {\bf 0}, \, \mu_m \geq  0, \nu_m \geq  0, \, m=1,2,\ldots,M. 
\end{aligned}
\end{equation*}
where ${\bf Z} \in \mathbb{S}^{DN}_{+}$ (we use the fact that $\mathbb{S}^{DN}_{+}$ is self-dual), ${\boldsymbol \mu} = [\mu_1,\mu_2,\ldots,\mu_M]^T \in \mathbb{R}^{M}$ and ${\boldsymbol \nu} = [\nu_1,\nu_2,\ldots,\nu_M]^T \in \mathbb{R}^{M}$ are the dual variables.
By eliminating $\nu_m$, the dual problem is simplified to
\begin{subequations}
\label{eq:dualapp1}
\begin{align} 
\hskip0.5mm &\argmax_{{\bf Z}, \, {\boldsymbol \mu}} \quad \lambda_{\rm eig}\,{\rm tr}\{{\bf Z}\}-{\bf 1}_M^T {\boldsymbol \mu}\label{eq:dualaapp1}\\
&\hskip1mm{\rm s.t.} \quad {\rm tr}\{{\bf F}_m{\bf Z}\}  \leq 1 + \mu_m, \, m=1,2,\ldots,M,  \label{eq:l1bapp1}\\ 
&\hskip10mm {\bf Z} \succeq {\bf 0}, \, \mu_m \geq  0, \, m=1,2,\ldots,M. \label{eq:dualcapp1}
\end{align}
\end{subequations}

\begin{figure*}[!t]
\normalsize 
\begin{equation}
\label{eq:Ldual}
\phi({\bf Z}, {\boldsymbol \mu}, {\boldsymbol \nu}) = \inf_{\bf w} {\mathcal L}({\bf w},{\bf Z}, {\boldsymbol \mu}, {\boldsymbol \nu}) =
  \begin{cases}
   - {\rm tr}\{-\lambda_{\rm eig}{\bf Z}\} -  {\bf 1}_M^T{\boldsymbol \mu} & \text{if }  {\rm tr}\{{\bf F}_m{\bf Z}\} + \nu_m = 1 + \mu_m, \, m=1,2,\ldots,M, \\
   -\infty       & \text{otherwise}.
  \end{cases}
\end{equation}
\hrulefill
\end{figure*}

\section{Projected Newton's method}\label{app:Newton's}
In order to analyze the complexity of the interior point methods, we briefly describe the projected Newton's method. The Newton's method for an SDP problem in the inequality form is adapted to suit our problem~\cite[Pg. 619]{Boyd}. 

The optimization problem in \eqref{eq:l1} can be approximated using the log-determinant barrier function which is given as
$$\argmin_{{\bf w} \in [0,1]^N} \quad  {\psi}({\bf w}) = t {\bf 1}_M^T{\bf w} - \ln {\rm det} \{\sum_{m=1}^{M} w_m {\bf F}_{m} - \lambda_{\rm eig} {\bf I}_{DN}\},$$
where $t >0$ is a parameter to tune the approximation. The projected Newton's update equation is given by
\begin{equation}
{\bf w}^{k+1} = \mathcal{P}_\mathcal{W}\left({\bf w}^{k} - \alpha^{k} \left({\frac{\partial^2 \psi({\bf w}^{k})}{\partial w_i^{k} \partial w_j^{k}}}\right)^{-1}\frac{\partial \psi({\bf w}^{k})}{\partial w_i^{k}}\right),
\end{equation}
where the entries of the Hessian matrix are given by
$$\left.\frac{\partial^2 \psi({\bf w})}{\partial w_i \partial w_j}\right\vert_{{\bf w}={\bf w}^{k}} = {\rm tr}\{{\bf S}^{-1}{\bf F}_i{\bf S}^{-1}{\bf F}_j\}, i,j=1,2,\ldots,M,$$ and the entries of the gradient vector are given by
$$\left.\frac{\partial \psi({\bf w})}{\partial w_i}\right\vert_{{\bf w}={\bf w}^{k}} = t + {\rm tr}\{{\bf S}^{-1}{\bf F}_i\}, i =1,2,\ldots,M.$$ Here, we have introduced the matrix ${\bf S} = \sum_{m=1}^{M} w_m {\bf F}_{m} - \lambda_{\rm eig} {\bf I}_{DN}$, and recall the projector operator $\mathcal{P}_{\mathcal{W}}(\cdot)$ defined in \eqref{eq:projection}. The step-length $\alpha^{k}$ is chosen by line-search.
\section{Power iterations for computing the minimum eigenvalue}\label{app:poweriter}
We briefly describe the power iterations~\cite{golub1996matrix} to compute the minimum eigenvalue of a matrix ${\bf F} \in \mathbb{S}^N$. Assuming ${\bf F}$ has ordered eigenvalues $\lambda_{\rm max} \geq \lambda_2 \cdots \geq \lambda_{\rm min}$, the power iterations
$$
{\bf v}^{k+1} = \frac{{\bf F}{\bf v}^{k}}{{\|{\bf F}{\bf v}^{k}\|}_2}, \, \text{and} \,\, \lambda^{k+1} = \frac{({\bf v}^{k+1})^T{\bf F}{\bf v}^{k+1}}{{\|{\bf v}^{k+1}\|}_2}, 
$$
converge to the eigenvector corresponding to the maximum eigenvalue ${\bf v}_{\rm max}$, and the maximum eigenvalue $\lambda_{\rm max}$, respectively, as $k \rightarrow \infty$. Here, we use ${\bf v}^0 = [1, {\bf 0}_{N-1}^T]^T$. By forming a matrix $\bar{\bf F} = \lambda_{\rm max}{\bf I}_N - {\bf F}$ which has the dominant eigenvalue $\lambda_{\rm max} - \lambda_{\rm min}$, we can apply the above power iterations on $\bar{\bf F}$ to compute $\lambda_{\rm max}- \lambda_{\rm min}$ and ${\bf v}_{\rm min}$, and thus the minimum eigenvalue of {\bf F} and it's corresponding eigenvector. 

\bibliographystyle{IEEEbib}
\bibliography{IEEEabrv,strings,refs}

\end{document}